\documentclass[iop,twocolumn,twocolappendix,numberedappendix]{emulateapj}
\usepackage{latexsym,graphicx,rotating,amsmath, epsfig, natbib}
\usepackage[usenames, dvipsnames]{color}

\shorttitle{Magnetic Wreaths and Cycles}
\shortauthors{Nelson et al.} 

\begin{document}

  \title{Magnetic Wreaths and Cycles in Convective Dynamos}

\author{Nicholas J. Nelson\altaffilmark{1}, Benjamin P. Brown\altaffilmark{2,3}, Allan Sacha Brun\altaffilmark{4}, Mark S. Miesch\altaffilmark{5}, \& Juri Toomre\altaffilmark{1} }
\affil{$^1$ JILA and Dept. of Astrophysical \& Planetary Sciences, University of Colorado, Boulder, CO 80309-0440}
\affil{$^2$ Dept. of Astronomy, University of Wisconsin, Madison, WI 53706-1582}
 \affil{$^3$ Center for Magnetic Self Organization in Laboratory and Astrophysical Plasmas, University of Wisconsin, 1150 University Avenue, Madison, WI 53706}
\affil{$^4$ Laboratoire AIM Paris-Saclay, CEA/Irfu Universit\'e Paris-Diderot CNRS/INSU, 91191 Gif-sur-Yvette, France.}
\affil{$^5$ High Altitude Observatory, NCAR, Boulder, CO 80307-3000}

\email{Contact Info:  nnelson@lcd.colorado.edu}

  \begin{abstract}
 Solar-type stars exhibit a rich variety of magnetic activity.  Seeking to explore the
convective origins of this activity, we have carried out a series of global 3D
magnetohydrodynamic (MHD) simulations with the anelastic spherical 
harmonic (ASH) code.  Here we report on the dynamo mechanisms achieved
as the effects of artificial diffusion are systematically decreased.  The simulations
are carried out at a nominal rotation rate of three times the solar value (3$\Omega_\odot$),
but similar dynamics may also apply to the Sun.  Our previous simulations 
demonstrated that convective dynamos can build persistent toroidal flux structures
(magnetic wreaths) in the midst of a turbulent convection zone and that high 
rotation rates promote the cyclic reversal of these wreaths.  Here we demonstrate
that magnetic cycles can also be achieved by reducing the diffusion, thus increasing
the Reynolds and magnetic Reynolds numbers.  In these more turbulent models, 
diffusive processes no longer play a significant role in the key dynamical balances that
establish and maintain the differential rotation and magnetic wreaths.  Magnetic
reversals are attributed to an imbalance in the poloidal magnetic induction by 
convective motions that is stabilized at higher diffusion levels.  Additionally, the 
enhanced levels of turbulence lead to greater intermittency in the toroidal 
magnetic wreaths, promoting the generation of  buoyant magnetic loops that 
rise from the deep interior to the upper regions of our simulated domain.  The
implications of such turbulence-induced magnetic buoyancy for solar and
stellar flux emergence are also discussed.
  \end{abstract}
\slugcomment{}


\section{Magnetic Variability in the Sun and in Sun-like Stars}

Informed and motivated by advances in observations of the Sun and solar analogues as well as the extensive theoretical framework of convective dynamo theory, we have undertaken a series of three-dimensional magnetohydrodynamic (MHD) simulations of convection and dynamo action in solar-type stars.  
These numerical experiments show that a rich variety of temporal variability in the magnetic topology, polarity, and field strength can be achieved without varying the rotation rate, even over the portion of parameter space accessible to our computational resources.  
We focus on models with rotation rates faster than the current solar rate $\Omega_\odot$, which accentuates the magnetic self-organization processes we are interested in exploring and which taps into the abundant observations of magnetic activity in young, rapidly-rotating, solar-like stars.

Our work builds upon three previous groups of simulations that show the same stellar configuration, namely considering dynamics within the deep convective envelope of our current sun, but having these nominally young stars rotate faster. 
\cite{Brown2008} began with hydrodynamic simulations involving a range of rotation rates up to $10 \Omega_\odot$, finding that strong differential rotation is realized, and that the columnar convection at low latitudes can exhibit significant modulation in amplitude with longitude, even appearing as nearly isolated active nests. 
\cite{Brown2010} examined dynamo action achieved in a MHD simulation carried out at $3 \Omega_\odot$, finding that the convection can build ordered global-scale magnetic fields that appear as two wreaths of strong toroidal field positioned above and below the equator.  
These striking structures can persist for long intervals despite being embedded within a turbulent convective layer.  
Turning to dynamo action proceeding at a faster rotation rate of $5 \Omega_\odot$, \cite{Brown2011} showed that self-consistently generated magnetic wreaths at low latitudes can undergo reversals in global magnetic polarity and even quasi-cycles of magnetic activity.  
The complex steps involved in the magnetic field reversals are accompanied by variations in the differential rotation, including bands of relatively fast and slow fluid propagating toward the poles.

As we decrease the dissipation in our simulations, we find that reversals of global magnetic polarity and cycles of magnetic activity are achieved.  
Despite more vigorous small-scale turbulence, these simulations still form global-scale magnetic wreaths in the bulk of their convective layers.  
Yet, the dynamical balances which maintain differential rotation and generate the mean toroidal magnetic field change as resolved turbulent transport assumes the dissipative role that was previously played by unresolved subgrid-scale (SGS) turbulent diffusion.  
We attribute the origin of magnetic cycles to a similar shift in the dynamical balance in the mean poloidal induction equation that had previously sustained steady wreaths.  
As the SGS dissipation is decreased, it is unable to offset the zonal component of turbulent electromotive force (EMF), which generates opposing poloidal field near the equator and thereby brings about the polarity reversal of the wreaths.  
A decrease in SGS dissipation also makes the wreaths more localized as coherent wreath segments over a limited range of longitudes as opposed to axisymmetric bands.  
This tends to decrease the mean field component while simultaneously increasing the field strength in the core of the wreaths.  
We argue that this has important implications for flux emergence, since it is these localized wreath cores that are most likely to trigger the magnetic buoyancy instability.

\subsection{Inspiration from Observational Advances}

As these simulations demonstrate, a rich variety of temporal variability can be achieved in dynamo models with only modest changes in their control parameters. 
Magnetic activity is a ubiquitous characteristic of sun-like stars and many stars exhibit cycles of magnetic activity.  
The best example is perhaps the sun's 22-year magnetic activity cycle. 
The interplay of turbulent convection, rotation, and stratification in the solar convection zone creates a cyclic dynamo which drives variations in the interior, on the surface, and throughout the sun's extended atmosphere \citep{Pinto2011}.  

The Sun is not alone in its magnetic variability.  
Solar-type stars generate magnetism almost without exception, particularly at rotation rates greater than that of the current sun.  
Young, rapidly rotating suns appear to have much stronger magnetic fields at their surfaces. 
Observations reveal a clear correlation between rotation and magnetic activity, as inferred from proxies such as X-ray and chromospheric emission \citep{Saar1999, Pizzolato2003, Wright2011}, however superimposed on this trend is considerable variation in the presence and period of magnetic activity cycles. 
There have been a number of attempts to monitor the magnetic activity cycles of other stars using solar-calibrated proxies for magnetic activity \citep[e.g.,][]{Baliunas1995, Hempelmann1996, Olah2009}. 
These programs require long, sustained periods of consistent observations, and are therefore rare.  
To date, the largest such project is the Mount Wilson HK survey, which measured chromospheric calcium lines as a proxy for magnetic activity for 111 solar-like stars over a 25-year period ending in 1991.  
In that study almost half of the stars showed cyclic behavior including 21 with regular periods between 7 and 25 years \citep{Baliunas1995}. 
The existence of sun-like stars without clear cycles of magnetic activity provide inspiration in this work for the study of a family of dynamo models that lie very closely together in parameter space but exhibit markedly different degrees of temporal variability in their large-scale magnetic features. 
Improved observational techniques including spot-tracking from Kepler photometry \citep{Meibom2011, Llama2012} and Zeeman-Doppler imaging \citep{Petit2008, Gaulme2010, Morgenthaler2012} may permit measurements of the size, frequency, and magnetic flux of starspots and the topology and spatial variability of photospheric magnetic fields. 
These developments are likely to provide new challenges to our existing theories of dynamo action and flux emergence in sun-like stars. 

Whatever the future of solar and stellar observations holds, one thing is clear: barring revolutionary advances in helio and asteroseismology, the information we have about solar and stellar magnetic activity is mainly limited to each star's surface and atmosphere.  
Furthermore, it is the magnetic flux and energy that passes through the solar surface that shapes the structure and evolution of the solar corona and heliosphere and regulates space weather.  
Thus, understanding the link between dynamo action in the interior and flux emergence at the surface is a vital area of research for solar and stellar physics.  
While the solar dynamo operates on a wide variety of scales in both size and time, our simulations seek to make contact with elements of the large-scale magnetic behavior on time-sales of years to decades.

\subsection{Building Upon Theoretical Frameworks}

Cyclic dynamos are fundamentally three-dimensional, nonlinear, and chaotic.  
In spite of this difficulty, much of the groundwork for modern dynamo theory has been laid in analytic mean-field models \citep[e.g.,][]{Parker1955, Moffatt1978, Krause1980}.  
The generation of toroidal field as differential rotation acts on a poloidal field, for example, can be well described using these models. 
The so-called $\Omega$-effect relies on shear from differential rotation in the convection zone or the tachocline at its base to stretch poloidal field into bands of toroidal field. 
The regeneration of poloidal field or the generation of opposite polarity poloidal field is parameterized in mean-field theory through the $\alpha$-effect. 
A number of theoretical dynamo models have been proposed, but as of yet no single numerical model has been able to capture all of the physical mechanisms required \citep[see review by][]{Charbonneau2010}.

To confront the complex nature of solar-like dynamo action, numerical models have been developed to explore aspects of various dynamo processes. 
Pioneering work by \cite{Gilman1983} and \cite{Glatzmaier1985} produced the first 3D MHD simulations of cyclic dynamo action in a rotating spherical shell.  
\cite{Miesch2006, Miesch2008} have explored the interplay of rotation, stratification, and moderately turbulent convection to produce strong differential rotation in a hydrodynamic setting. 
When magnetism is added in this context the resulting dynamo produces reversals of global polarity but is dominated by non-axisymmetric fields with little global organization \citep{Brun2004}. 
By adding an overshooting region of strong shear which mimics the solar tachocline, global-scale organization of the toroidal field was achieved by, but without reversals in global magnetic polarity over about 30 simulated years \cite{Browning2006}.  
Recent work by \cite{Ghizaru2010} has shown large-scale organization of the toroidal field as well as magnetic activity cycles in a solar-like simulation; regular reversals of global magnetic polarity with a roughly 60 year period for a complete cycle were achieved.  
Further work by \cite{Racine2011} has interpreted these results in terms of mean-field dynamo theory.

\begin{deluxetable*}{lcccccccccccccc}
  \tabletypesize{\footnotesize}
    \tablecolumns{14}
    \tablewidth{0pt}  
    \tablecaption{Overview of Dynamo Cases
    \label{table:dynamo_sim_parameters}}
    \tablehead{\colhead{Case}  &  
      \colhead{$N_r,N_\theta,N_\phi$} &
      \colhead{Ra} &
      \colhead{Ta} &
      \colhead{Re} &
      \colhead{Re$'$} &
      \colhead{Rm} &
      \colhead{Rm$'$} &
      \colhead{Ro} &
      \colhead{Roc} &
      \colhead{$\nu$} &
      \colhead{$\eta$} &
      \colhead{Pm} &
     \colhead{$T_E$}
   }
   \startdata
  D3    & $ \phn 97 \times  256 \times  512\phn$ &    3.28$ \times 10^{  5}$ &     1.22$ \times 10^{  7}$ & 173 &  104 &   86 &   52 &    0.374 &    0.315 &     13.2 &     26.4 & 0.5&  61.6 \\[3mm]
  D3a   & $ \phn 97 \times  256 \times  512\phn$ &    5.84$ \times 10^{  5}$ &     2.41$ \times 10^{  7}$ & 244 &  154 &  122 &   77 &    0.447 &    0.295 &    9.40 &     18.8 & 0.5& 67.1  \\
  D3b   & $ 145 \times  512 \times 1024$ &    1.11$ \times 10^{  6}$ &     6.08$ \times 10^{  7}$ & 343 &  273 &  171 &  136 &    0.566 &    0.257 &    5.92 &     11.8 &  0.5 & 16.9 \\[3mm]
  D3-pm1 & $ 145 \times  256 \times  512\phn$ &    2.98$ \times 10^{  5}$ &     1.22$ \times 10^{  7}$ & 149 &  102 &  149 &  102 &    0.372 &    0.300 &     13.2 &     13.2 &  1 & 18.8 \\
  D3-pm2 & $ 145 \times  512 \times 1024$ &    3.08$ \times 10^{  5}$ &     1.22$ \times 10^{  7}$ & 145 &  101 &  291 &  202 &    0.370 &    0.306 &     13.2 &    6.60 &  2 & 13.6 \\ [3mm]
S3 &  $145 \times 512 \times 1024$  &  $7.68 \times 10^8$ & $4.46 \times 10^{10}$ & 8050 &  5750  & 4030 & 2880 & 0.581 & 0.262 & 0.218 & 0.435 & 0.5 & 4.01
 \enddata
\tablecomments{Dynamo simulations at three times the solar rotation rate.
        All simulations have inner radius 
	$r_\mathrm{bot} = 5.0 \times 10^{10}$cm and outer radius of 
        $r_\mathrm{top} = 6.72 \times 10^{10}$cm, with 
	$L = (r_\mathrm{top}-r_\mathrm{bot}) = 1.72 \times 10^{10}$cm
	the thickness of the spherical shell.
	Evaluated at mid-depth are the
	Rayleigh number $\mathrm{Ra} = (-\partial \rho / \partial S)
	(\mathrm{d}\bar{S}/\mathrm{d}r) g L^4/\rho \nu \kappa$, 
	the Taylor number $\mathrm{Ta} = 4 \Omega_0^2 L^4 / \nu^2$, 
	the rms Reynolds number $\mathrm{Re}  = v_\mathrm{rms} L /\nu$ and
	fluctuating Reynolds number $\mathrm{Re}' = v_\mathrm{rms}' L /\nu$,
	the magnetic Reynolds number $\mathrm{Rm} = v_\mathrm{rms} L/\eta$ 
	and fluctuating magnetic Reynolds number 
        $\mathrm{Rm}' = v_\mathrm{rms}' L/\eta$, 
	the Rossby number $\mathrm{Ro} = \omega / 2 \Omega_0$,
	and the convective Rossby number 
	$\mathrm{Roc} = (\mathrm{Ra}/\mathrm{Ta} \, \mathrm{Pr})^{1/2}$.
	Here the fluctuating velocity $v'$ has the axisymmetric
        component removed: $v' = v - \langle v \rangle$, 
        with angle brackets denoting an average in longitude.
	For all simulations, the Prandtl number $\mathrm{Pr} = \nu / \kappa$ is 0.25 
	and the magnetic Prandtl number $\mathrm{Pm}=\nu/\eta$ ranges between 0.5 and 4.  
	The viscous and magnetic diffusivity, $\nu$ and $\eta$, are
	quoted at mid-depth (in units of $10^{11}~\mathrm{cm}^2 \; \mathrm{s}^{-1}$).
         The total evolution time $T_E$ for each simulation is given in years.  
         The values for case S3 with the dynamic Smagorinsky SGS model utilize the mean viscosity at mid-convection zone averaged on horizontal surfaces as well as in time.  
         For case S3 using the dynamic Smagorinsky SGS model, the values quoted are based on the time-averaged rms viscosity, conductivity, and resistivity at mid-depth, noting that these diffusion coefficients have near hundred-fold spatial variations. }
\end{deluxetable*} 

Additional insights into the solar dynamo have been realized through local simulations that do not model the full spherical geometry in order to achieve higher resolution in a local domain \citep{Cline2003, Vasil2009}.  
The recent work of \cite{Guerrero2011} has shown that dynamo action in a domain with convection and a forced shear layer can produce strong dynamo action and yield buoyant magnetic structures.  
Another approach using helical forcing in a Cartesian domain has shown that large-scale magnetic structures which undergo regular reversals in polarity can be achieved even without convection or rotation, hinting at the key role of helicity \citep{Mitra2010}.

Here we expand upon the work of \cite{Brown2010, Brown2011} in exploring global 3D simulations of dynamo action in sun-like stars rotating faster than the current solar rotation rate.  
We report on a suite of simulations at $3 \Omega_\odot$ which explore the transition from dynamos with persistent toroidal fields to cyclic dynamos by decreasing the level of explicit diffusion in the simulations.  
Although the simulations reported here are ostensibly at a rotation rate of $3 \Omega_\odot$, it is important to realize that the dynamics we describe may not be restricted to young stars. 
With regard to the generation of magnetic wreaths, the most salient nondimensional parameter of the physical system is the Rossby number, Ro$\, = \omega / (2\Omega_0)$, which is small in both the Sun and these models.
In this series of models, we begin to address the feasibility of the wreath-building dynamo mechanisms at higher levels of turbulence and study the global-scale reversals and cycles of magnetic activity achieved, which will be the focus of \S 3-6.  We also begin to explore the subtle but crucial link between wreath-building convective dynamos and flux emergence, which will be the focus of \S 7.

\section{Dynamos at $3 \Omega_\odot$}

\begin{figure*}
\begin{center}
  \includegraphics[width=0.95\linewidth]{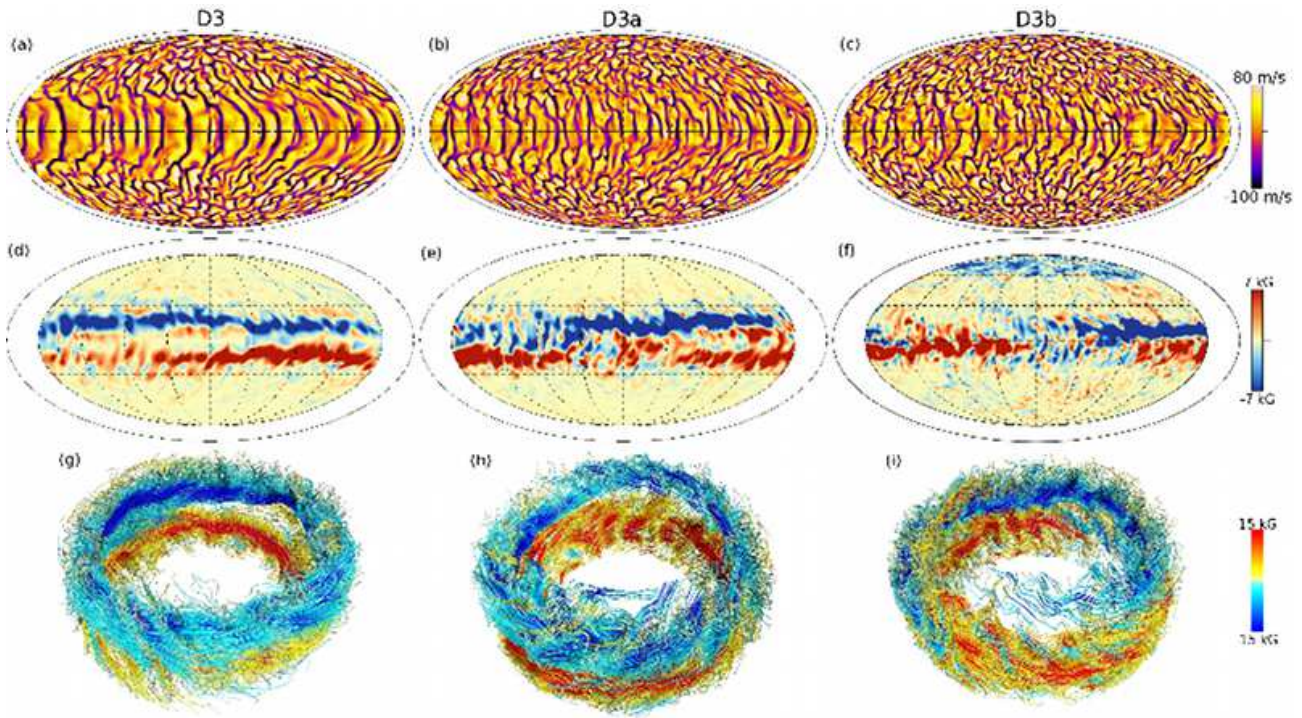}
  \caption{Magnetic wreaths.  (a)-(c) Shown in global Mollweide view (equator at middle, poles at top and bottom) is the radial velocity of the convection at $0.95 R_\odot$ in cases D3, D3a, and D3b, respectively.  (d)-(f) Also in Mollweide few, longitudinal magnetic field $B_\phi$ at mid-convection zone at times $t_1$ indicated in Figure~\ref{fig:TimeLat Stack}.  (g)-(i) Shown at the same times for each case is a 3D field line rendering of the magnetic wreaths near the equator.  In both types of display for the magnetic field, color gives the polarity and amplitude of the longitudinal field (red positive, blue negative).  Times shown correspond to $t_1$ for each case in Figure~\ref{fig:TimeLat Stack}.
  \label{fig:meet D3 family}}
  \end{center}
\end{figure*}

We study convection and dynamo action in the deep interior of solar-like stars using the anelastic spherical harmonic (ASH) code \citep{Clune1999, Brun2004}.  
Our simulation approach is briefly described here, but is more fully explained in \cite{Brown2010}.  
ASH solves the anelastic MHD equations in a rotating spherical shell with a background stratification taken from a 1D model of solar structure. 
We focus on simulating the bulk of the solar convection zone from $0.72 R_\odot$ to $0.97 R_\odot$ ($R_\odot$ is solar radius) with a density contrast of about 25.  
We do not model the near-surface layers of the sun, for we are limited by the anelastic approximation to subsonic flows.  
Additionally we cannot resolve the small-scales of motion needed to simulate granular and supergranular scales.  
We also do not include the stably-stratified radiative zone or the tachocline in these simulations, although simulations including those components are an active area of research \citep[see][]{Brun2011}.  
We have done some preliminary work in adding a tachocline to these simulations and found that it does not drastically change the dynamo action in the bulk of the convective layer. 
The effects of a tachocline will will explored further in a future paper.  
Our results tend to support the recent studies with mean-field dynamo models, which suggests that the differential rotation of the convection zone may play a greater role in the generation of toroidal magnetic field than the tachocline \cite[e.g.,][]{Dikpati2006, Munoz2009}. 
We use impenetrable and stress-free boundary conditions on both the top and bottom of the domain. 
We impose the entropy gradient at the top and bottom of the domain for the thermal boundary conditions.  
For the magnetic fields we use a perfect conductor condition on the bottom boundary and match to an external potential field on the top boundary.  
These conditions and our evolution equations are described in detail in \cite{Brown2010}.

ASH is a large-eddy simulation which employs a subgrid-scale model to account for the effects of unresolved scales of motion.  
The standard subgrid-scale (SGS) model in ASH simulations uses enhanced values of viscosity, thermal diffusivity, and magnetic resistivity relative to those expected from atomic values in order to represent additional mixing due to unresolved turbulent motions. 
In this enhanced eddy SGS model, viscosity $\nu$, thermal diffusivity $\kappa$, and magnetic resistivity $\eta$ all scale as $\bar{\rho}^{-1/2}$, where $\bar{\rho}$ is the spherically-symmetric background density of the simulation.  
This prescription, along with constant Prandtl and magnetic Prandtl numbers throughout the domain, follows that of \cite{Brown2010, Brown2011}.  
All cases presented in this paper use Pr = $\nu /  \kappa = 0.25$, but variable Pm (see Table~\ref{table:dynamo_sim_parameters}). 

In addition, we have also implemented a more complex SGS treatment, the dynamic Smagorinsky model developed by \cite{Germano1991}.
By using the dynamic Smagorinsky model in ASH simulations we are able to reduce the mean diffusion at mid-convection zone by a factor of 50 without an increase in resolution. 
Our implementation of the dynamic Smagorinsky model is summarized in Appendix A.  
This SGS treatment is only used in case S3, which was first presented in \cite{Nelson2011}.

Table~\ref{table:dynamo_sim_parameters} presents the computational resolution, relevant non-dimensional parameters, diffusion coefficients, and total evolution time for each of the six cases we will discuss here.  
We have explored two main branches in parameter space.  
The first branch includes cases D3, D3a, and D3b, where viscosity $\nu$, thermal diffusivity $\kappa$, and magnetic resistivity $\eta$ have all been dropped together, thus keeping a constant magnetic Prandtl number.  
The second branch includes cases D3, D3-pm1, and D3-pm2, where $\nu$ and $\kappa$ are held constant and only $\eta$ is decreased, resulting in increasing magnetic Prandtl numbers.  
We will refer to the two branches as the constant Pm and increasing Pm branches, respectively.  
The constant Pm branch was found to be more compelling, as cases D3a and D3b generally produced strong magnetic wreaths that were anti-symmetric about the equator, whereas the high Pm branch produced a wider variety of symmetric and anti-symmetric toroidal field states and was therefore less amenable to study.  
Such behavior is not unexpected as dynamos with higher magnetic Prandtl number tend to promote small-scale dynamo action.  We will generally focus on the constant Pm branch of simulations while referencing the increasing Pm branch to provide additional insight.

\begin{deluxetable*}{lcccccccccc}
  \tabletypesize{\footnotesize}
    \tablecolumns{9}
    \tablewidth{0pt}  
    \tablecaption{Volume-Averaged Energy Densities and Differential Rotation Rates
    \label{table:dynamo energies}}
    \tablehead{\colhead{Case}  &  
      \colhead{Total ME} &
      \colhead{TME} &
      \colhead{PME} &
      \colhead{FME} &
      \colhead{Total KE} &
      \colhead{DRKE} &
      \colhead{MCKE} &
      \colhead{FKE} &
             \colhead{$\Delta \Omega_r$} &
       \colhead{$\Delta \Omega_\theta$}
   }
   \startdata
  D3 & 0.68 (9\%) & 0.29 (43\%) & 0.029 (4\%) & 0.36 (53\%) & 6.67 (91\%) & 4.35 (65\%) & 0.010 (0.1\%) & 2.31(35\%) & 112 & 192 \\[3mm]
  D3a & 0.88 (12\%) & 0.32 (36\%) & 0.030 (3\%) & 0.52 (59\%) & 6.41 (88\%) & 3.71 (58\%) & 0.011 (0.2\%) & 2.68 (42\%) & 101 & 163 \\
  D3b & 0.82 (13\%) & 0.10 (12\%) & 0.011 (1\%) & 0.70 (85\%) & 5.42 (87\%) & 2.45 (45\%) & 0.012 (0.2\%) & 2.96 (55\%) & 95 & 131 \\[3mm]
  D3-pm1 & 1.04 (18\%) & 0.26 (25\%) & 0.033 (3\%) & 0.75 (72\%) & 4.87 (82\%) & 2.63 (54\%) & 0.010 (0.2\%) & 2.23 (46\%) & 87 & 139 \\
  D3-pm2 & 1.17 (21\%) & 0.15 (13\%) & 0.028 (2\%) & 0.99 (85\%) & 4.34 (75\%) & 2.29 (53\%) & 0.009 (0.2\%) & 2.04 (47\%) & 74 & 121 \\[3mm]
S3 & 0.83 (13\%) & 0.072 (9\%) & 0.0060 (0.7\%) & 0.75 (91\%) & 5.50 (87\%) & 2.32 (42\%) & 0.013 (0.2\%) & 3.17 (58\%) & 95 & 133 
\enddata
\tablecomments{Volume-averaged magnetic and kinetic energies for dynamo simulations at three times the solar rotation rate, as well as the magnitude of the differential rotation contrast in radius at the equator $\Delta \Omega_r$ and the average contrast at the top of the simulated domain between the equator and $\pm 60^\circ$ latitude $\Delta \Omega_\theta$.  Shown in units of $10^{6}$ erg cm$^{-3}$ are the total magnetic energy (Total ME), axisymmetric toroidal magnetic energy (TME), axisymmetric poloidal magnetic energy (PME), fluctuating magnetic energy (FME), total kinetic energy (Total KE), differential rotation kinetic energy (DRKE), meridional circulation kinetic energy (MCKE), and fluctuating kinetic energy (FKE). The percentage of the total energy is shown for total magnetic energy (Total ME) and total kinetic energy (Total KE). The percentage of the total magnetic or kinetic energy for each component is shown in parentheses. Values for differential rotation rates are in units of nHz ($3 \Omega_\odot = 1240$ nHz). Values are averaged in time over long intervals.}
\end{deluxetable*}

Case D3 was initiated from a well developed hydrodynamical simulation that was seeded with a small random magnetic field.  
Each subsequent case along both branches was started from the preceding case.  
Thus both cases D3a and D3-pm1 were started using case D3 as initial conditions, case D3b was started using case D3a, and so on.  
We have re-started case D3a from a random seed field to verify that it settles into a similar region of solution space as the version continued from case D3 and found no strong differences in the time-averaged behavior over several thousand days.

\section{Magnetic wreaths}

The dominant magnetic structures built by each of these simulations are the low latitude bands of predominately toroidal field, which we term wreaths.  
These wreaths are generally anti-symmetric about the equator, though symmetric states are observed along with states where one hemisphere displays a wreath while the other does not.  
These irregular states are most common along the increasing Pm branch of our simulations.  
The wreaths in case D3 are discussed extensively by \cite{Brown2010} and additional wreaths are analyzed at somewhat faster rotation rate ($5 \Omega_\odot$) by \cite{Brown2011}.

\subsection{Magnetic Topology}

Figure~\ref{fig:meet D3 family} shows snapshots of the turbulent convection and the wreaths for cases D3, D3a, and D3b at mid-convection zone in global Mollweide view as well as at low latitudes in a 3D volume rendering of magnetic field lines colored by $B_\phi$.  
In all three cases strong longitude-averaged fields are present at low latitudes, however the nature of the wreaths change from case D3 where axisymmetric fields dominate to case D3b where a significant axisymmetric field component is present but not dominant.  
In case D3b the morphology has changed such that the wreaths are confined in longitudinal extent.  
Figure ~\ref{fig:meet D3 family} shows a typical field configuration, but the wreaths are observed at various times to extend over as little at 45 degrees and as much as 270 degrees in longitude.  
All three cases show extensive connectivity between the wreaths and the surrounding domain where magnetic fields are moderate in strength but far less coherent.  
The wreaths are strongly modulated by the convective flows, producing a ragged appearance that is particularly noticeable in case D3b but present in all three cases.  
In the more turbulent cases there are also significant small-scale magnetic fields at moderate to high latitudes, and occasional locally-generated wreath-like structures near the poles which persist for less than about 100 days at a time.

The shift from structures dominated by axisymmetric fields in case D3 to the patchy wreaths in case D3b is illustrated by the changes in the time- and volume-averaged energy densities shown in Table~\ref{table:dynamo energies}.  
Between cases D3 and D3b there is a roughly 30\% increase in the total magnetic energy of the simulation, though the energy in both the axisymmetric toroidal (TME) and poloidal (PME) fields decreases by roughly a factor of three.  
The doubling of the energy in the non-axisymmetric magnetic fields more than compensates for the decrease in mean fields.  
When compared with the kinetic energy densities, the changes in the magnetic energies becomes even more striking. 
Viscous, thermal, and magnetic diffusion in case D3b are all reduced by the same factor relative to case D3. 
However the total kinetic energy in case D3b dropped by 19\%. 
The non-axisymmetric kinetic energy (FKE) rose only moderately compared to the decrease in differential rotation kinetic energy (DRKE).
The high magnetic Prandtl number cases also show a tendency towards larger total and fluctuating magnetic energies, as well as reduced axisymmetric toroidal magnetic energy as the magnetic diffusion is reduced. 

It is illustrative to compare cases D3b and D3-pm1, as they have roughly equal levels of magnetic diffusion, with case D3b having comparatively lower levels of viscosity and thermal diffusion.  
The largest differences are in the axisymmetric magnetic energies which are both about three times greater in case D3-pm1 than in case D3b.  
This may be due to the more laminar flow in case D3-pm1, which would tend to create fewer sharp gradients in the large-scale magnetic structures and thus lower the effective dissipation in case D3-pm1 compared to case D3b, even though the diffusion coefficients in the induction equation are nearly the same.
Case D3-pm1 also show significantly less differential rotation contrast both in radius and latitude compared to case D3b, pointing to the key role of magnetic torques in decreasing differential rotation, which will be discussed further in \S5.

\begin{figure}
\begin{center}
  \includegraphics[width=0.95\linewidth]{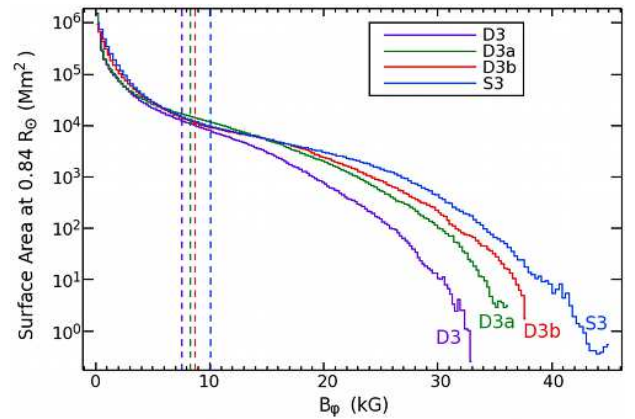}
  \caption{Probability distribution functions for unsigned $B_\phi$ at mid-convection zone for cases D3 (purple), D3a (green), D3b (red), and S3 (blue) showing the surface area covered by fields of a given magnitude. Distributions are averaged over about 300 days when fields are strong and as steady as possible. Dashed vertical lines show the field-strength at which equipartition is achieved with the maximum fluctuating kinetic energy (FKE) at mid-convection zone for each case. Case D3b shows a deficit of field in the $10$ kG range, but an excess of surface area covered by extremely strong fields above $25$ kG range, as well as higher peak field strengths. Case S3 shows significantly greater regions of fields in excess of 20 kG than all other cases.
  \label{fig:Bphi PDFs}}
  \end{center}
\end{figure}

\subsection{Non-axisymmetric Fields}

Our discussion of the magnetic wreaths to this point has focused on the axisymmetric fields, which are progressively weaker in moving from case D3 to case D3b. 
While the axisymmetric fields weaken with increased turbulence, very strong fields become more common when measured by the fraction of the domain they occupy. 
Figure~\ref{fig:Bphi PDFs} shows the probability distribution function for $B_\phi$ at mid-convection zone in cases D3, D3a, D3b, and S3.
While case D3b has a deficit of fields around 10 kG compared to case D3a, there is a clear excess of fields above 20 kG.  
Interestingly the distribution for case D3b is greater than that for case D3 for all but the smallest bin, indicating that while case D3 may have stronger axisymmetric fields in the low latitude wreaths, case D3b compensates by having higher amplitude fluctuating fields throughout the domain. 
The peak field strength at mid-convection zone is 32 kG in case D3, 36 kG in case D3a, and 38 kG in case D3b.  
Near the base of the convection zone case D3b exhibits even stronger fields of up to 44 kG. 
Case S3 posses magnetic fields of up to 45 kG at mid-convection zone and 52 kG near the base of the convective layer.
For all four cases fields are seen well in excess of equipartition energies with the maximum fluctuating kinetic energy of the flows.
This is a clear indication of turbulent intermittency in the magnetic fields.

A statistical measure of turbulent intermittency is the time-averaged excess kurtosis given by 
\begin{equation}
  \operatorname{kurt} \{ B_\phi \} = \frac{ \int^\infty_{-\infty} \left( B'_\phi - \bar{B}_\phi \right)^4 f \left( B'_\phi \right) dB'_\phi }{ \left[ \int^\infty_{-\infty}  \left( B'_\phi - \bar{B}_\phi \right)^2 f \left( B'_\phi \right) dB'_\phi \right]^2 } - 3 ,
\end{equation} 
where $ f ( B'_\phi )$ is the probability distribution function \citep[see][]{Pope2000}.  
For reference a Gaussian distribution would have an excess kurtosis of 0.  
The level of turbulent intermittency is measured by how leptokurtic the distribution is found to be, with large values corresponding to increased intermittency. 
For case D3 $\operatorname{kurt} \{ B_\phi \} = 9.6$, while for case D3a $\operatorname{kurt} \{ B_\phi \} = 10.5$, and for case D3b $\operatorname{kurt} \{ B_\phi \} = 12.1$.  
Leptokurtic distributions are likely to experience strong coherent structures, such as the strong regions of coherent toroidal field in these simulations. 
At even lower levels of diffusion than can be realized with the enhanced eddy SGS model, the strong-field regions become sufficiently buoyant and coherent so as to form buoyant magnetic loops as realized in case S3 \citep{Nelson2011}, for which $\operatorname{kurt} \{ B_\phi \} = 15.6$.
Highly leptokurtic distributions like these indicate that extreme events are enhanced relative to a Gaussian distribution, and the trend towards increasing kurtosis as simulations become more turbulent points to turbulent amplification of magnetic fields.
As we will discuss further in \S 7, this provides an alternate pathway to produce regions within the larger wreaths which can be amplified through turbulent intermittency to produce coherent regions of strong magnetic field, which can then become buoyant. 
We term this the turbulence-enabled magnetic buoyancy paradigm.

\begin{figure}
\begin{center}
  \includegraphics[width=\linewidth]{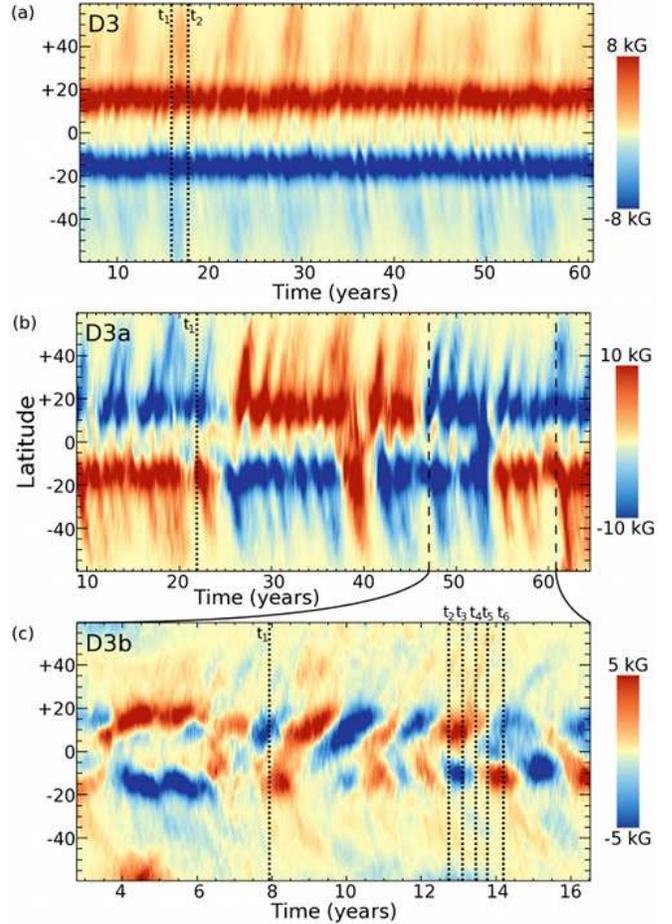}
  \caption{Time-latitude plots of longitude averaged toroidal magnetic field $\langle B_\phi \rangle$ at $0.79 R_\odot$ for (a) case D3 over about 56 years, (b) case D3a over the same amount of time, and (c) case D3b over about 13 years. Dotted lines show times referenced in Figures~\ref{fig:meet D3 family}, \ref{fig:D3b reversal}, \ref{fig:Aphi Balance D3}, \ref{fig:Aphi Balance D3b},  and~\ref{fig:Time Spec}. Dashed lines on (b) indicate the time period covered by (c). Case D3b was started from case D3a at $t_2$ (dotted line). The evolution of case D3b is limited by the increased computational cost of the higher resolution required for computational stability.
  \label{fig:TimeLat Stack}}
\end{center}
\end{figure}

\section{Cyclic Reversals Achieved by Reducing Diffusion}  

In addition to building strong magnetic wreaths, cases D3a and D3b exhibit cyclic reversals of global magnetic polarity.  
As is believed to occur in the sun, the general pattern of the cycles is that the toroidal fields peak at roughly the time when the poloidal field is reversing sign, and the poloidal fields peak in amplitude when the toroidal fields are reversing sign. 
There are also a number of variations on this pattern, where one hemisphere may develop considerably stronger fields than the other or where both hemispheres have the same sense of toroidal field, pointing to large contributions at these times from quadripolar poloidal fields.  
Cases D3-pm1 and D3-pm2 also display strong variations in the strength and topology of their axisymmetric fields. 
However the irregularities are more pronounced for these cases over the time simulated.

\begin{figure}
\begin{center}
  \includegraphics[width=0.95\linewidth]{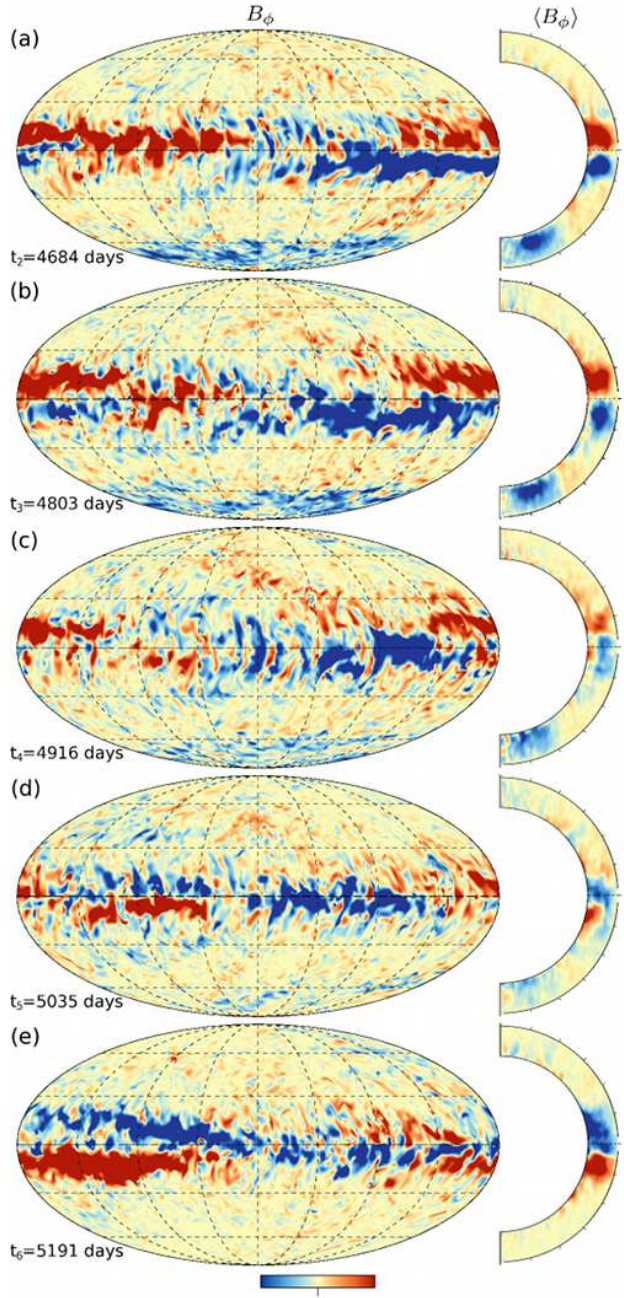}
  \caption{Reversal in magnetic polarity of the toroidal wreaths in case D3b shown in $B_\phi$ in Mollweide projection at mid-convection zone on left, and in $\langle B_\phi \rangle$ in longitudinal average over latitude and radius on right. Color indicates strength of toroidal magnetic field with the color table saturating at $\pm 7$ kG for the Mollweide images and $\pm 3$ kG for the longitudinal averages. Five snapshots corresponding to $t_2$ through $t_6$ from Figure~\ref{fig:TimeLat Stack}(c) are shown each separated by roughly 120 days. 
  \label{fig:D3b reversal}}
\end{center}
\end{figure}

\begin{figure}
\begin{center}
 \includegraphics[width=0.95\linewidth]{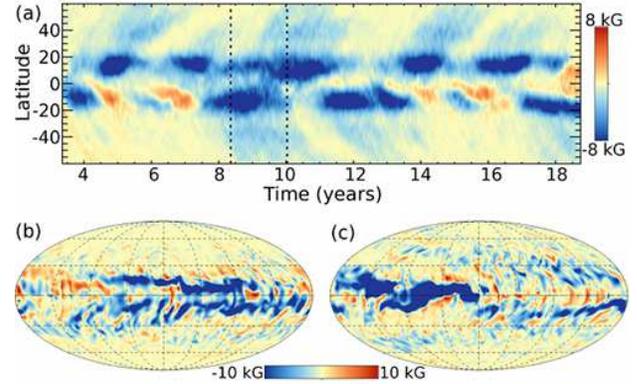}
 \caption{(a)~Time evolution in case D3-pm1 of the axisymmetric toroidal magnetic field at 0.79 $R_\odot$ over roughly 15 years of simulated time.  Strong variability of the mean fields is seen in both hemispheres. (b, c)~Companion snapshots of $B_\phi$ at 0.84 $R_\odot$ showing the spatial variability and non-axisymmetric nature of the wreaths.  Successive snapshot times are indicated by dashed lines in (a).
\label{fig:Pm1}}
\end{center}
\end{figure}

\subsection{Reversals in Global Magnetic Polarity}

Figure~\ref{fig:TimeLat Stack} shows the temporal evolution of the longitude-averaged toroidal field $\langle B_\phi \rangle$ at mid-convection zone over the history of cases D3 (Figure~\ref{fig:TimeLat Stack}(a)), D3a (Figure~\ref{fig:TimeLat Stack}(b)), and D3b (Figure~\ref{fig:TimeLat Stack}(c)).  
In case D3 we see persistent wreaths centered at about $20^\circ$ above and below the equator.  
These wreaths persist for about 68 years or as long as we have run the simulation.  
The polarity of the wreaths is constant in time, though variations on roughly 6 year time scales can be seen in both the amplitude of the low latitude wreaths as well as the propagation of field to higher latitudes.  
The behavior of this case is discussed in detail in \cite{Brown2010}. 
Figure~\ref{fig:TimeLat Stack}(b) shows case D3a over a comparable length of time as in the first panel.  
Case D3a undergoes reversals in global magnetic polarity as well as three significant irregular states.  
Additionally there are modulations in the amplitude of the wreaths and poleward movements of field on roughly 3 year time scales. 
These variations are not always synchronized between the two hemispheres, and neither are the reversals, indicating that the poloidal field can have a complicated structure.

Figure~\ref{fig:TimeLat Stack}(c) shows the temporal evolution of $\langle B_\phi \rangle$ at mid-convection zone for case D3b over about 13 years, with indications of cycles of magnetic activity and reversals of global polarity.  
We have simulated 10 reversals as measured by the time-smoothed antisymmetric component of $\langle B_\phi \rangle$ changing sign. 
The time between reversals ranges from 0.6 to 1.9 years and, as in case D3a, the two hemispheres are not always synchronized.  
There are several times when one hemisphere shows significantly stronger fields than the other or when both hemispheres have the same sense of fields.  
This is partly due to the averaging procedure used to create these figures and the fact that we are only looking at a single depth.  
Analysis of the full 3D data shows that there is almost always a wreath-like structure in each hemisphere.

Figure~\ref{fig:D3b reversal} shows a sequence of snapshots of $B_\phi$ at mid-convection zone in case D3b over a full spherical shell and of $\langle B_\phi \rangle$ over the domain before, during, and after a reversal in the polarity of the wreaths.  
Each snapshot is roughly 120 days after the previous snapshot.  
The wreaths start appearing as strong mean-field structures in the longitudinal average, but the non-averaged cut at mid-convection zone shows that there is significant longitudinal variation in the wreaths, with the northern wreath covering roughly $120^\circ$ and the southern wreath covering $180^\circ$ in longitude.  
There is also substantial evidence for interactions between the wreaths near the center of the image in Figure~\ref{fig:D3b reversal}(a).  
As time progresses, the axisymmetric fields weaken as the non-axisymmetric components begin to dominate.  
Small patches of strong field persist, but they are largely washed out in the longitudinal averages.  
After about 480 days (Figure~\ref{fig:D3b reversal}{d}) strong patches of opposite polarity field begin to appear and by the final frame (Figure~\ref{fig:D3b reversal}{e}) the strong mean fields have been reestablished in the opposite hemispheres from the initial configuration.

\subsection{Variability at Higher Magnetic Prandtl Number}

The simulations on the increasing Pm branch also show increased temporal variability relative to case D3. 
There is also evidence for a change in the nature of the dynamo action in these simulations.  
Figure~\ref{fig:Pm1} shows the evolution of $\langle B_\phi \rangle$ at 0.79 $R_\odot$ over the history of case D3-pm1, along with snapshots of the toroidal magnetic field at mid-convection zone at three representative times.  
This case has selected a configuration of toroidal field that is largely symmetric about the equator and of essentially the same polarity at most times. 
Some periods of positive polarity field are seen, though the dominant field in both hemispheres is clearly of negative polarity.  
Unlike cases D3a and D3b, case D3-pm1 does not undergo a true global reversal of magnetic polarity.  
It does, however, exhibit strong temporal variability in the wreaths seen in both hemispheres to an extent not seen in cases D3 or D3a. 

\begin{figure}
\begin{center}
 \includegraphics[width=0.95\linewidth]{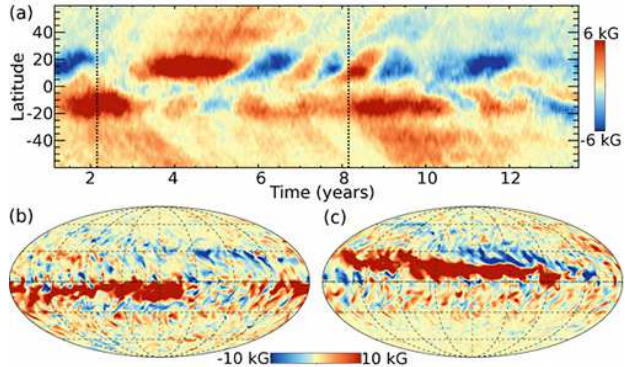}
 \caption{(a)~Time evolution in case D3-pm2 of the axisymmetric toroidal magnetic field at 0.79 $R_\odot$ over roughly 13 years of simulated time.  Strong variability of the mean fields is seen in both hemispheres, along with irregular reversals in polarity, at times in only one hemisphere and at other times globally. (b, c)~Companion snapshots of $B_\phi$ at 0.84 $R_\odot$ showing the spatial variability and non-axisymmetric nature of the wreaths.  Successive snapshot times are indicated by dashed lines in (a).
\label{fig:Pm2}}
\end{center}
\end{figure}

Figure~\ref{fig:Pm2} shows a similar view of case D3-pm2 over its temporal evolution.  
Again it tends to avoid the anti-symmetric states characteristic of cases D3, D3a, and, to a lesser extent, D3b.  
This case, however, does exhibit clear reversals of global magnetic polarity.  
Interestingly, these reversals do not appear to occur at regular intervals, and often one hemisphere can reverse without a noticeable change in the other hemisphere.  
As an example, the southern hemisphere maintains a positive polarity wreath between about $t = 5.5$ years and $t = 10.5$ years while the northern hemisphere exhibits four reversals in that same time interval.

The preference for irregular polarity states in $B_\phi$ along the increasing Pm branch is clearly related to the decreased level of magnetic diffusion, though it may also be indicative of a shift in behavior due to the transition from small to large magnetic Prandtl number.  
In cases D3, D3a, and D3b magnetic diffusion occurs on scales larger than those related to the diffusion of momentum.  
This tends to promote the concentration of magnetic energy at large scales.  
For high Pm dynamos, the resistive scale is smaller than the viscous scale, which tends to promote the growth of magnetic energy at small scales \citep[e.g.,][]{Schekochihin2004}. 
There is still considerable large-scale organization of magnetic field by the differential rotation, but the increasing Pm branch exhibits less ordered behavior than the constant Pm branch of simulations.

When examining the relative importance of decreased magnetic diffusion and increased levels of turbulence, it is perhaps most instructive to compare cases D3b and D3-pm2.  
Table~\ref{table:dynamo energies} shows that the division of magnetic energies between the axisymmetric toroidal, axisymmetric poloidal, and fluctuating magnetic fields are roughly equivalent in the two cases, although case D3-pm2 has more magnetic energy overall.  
The kinetic energies in case D3-pm2 are, however, more similar to case D3 than case D3b with the exception of decreased differential rotation kinetic energy due to enhanced Lorentz force feedbacks.  
This suggests that the onset of reversals is driven primarily by decreasing magnetic diffusion rather than by some subtle shift in the velocity fields or correlations between magnetic fields and velocities on small scales.

\begin{figure*}
\begin{center}
 \includegraphics[width=0.95\linewidth]{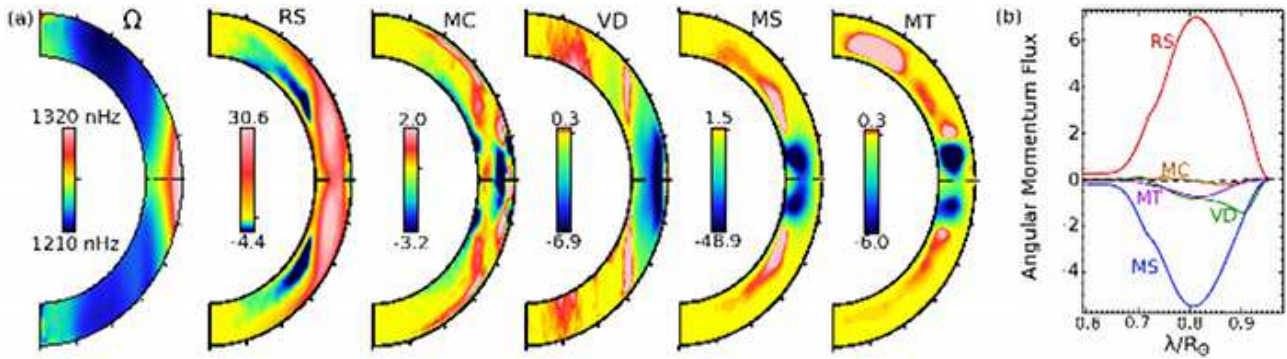}
 \caption{Differential rotation and the terms contributing to the accompanying redistribution of angular momentum in case D3b averaged over many magnetic cycles. (a) Angular velocity $\Omega$ profile with radius and latitude, accompanied in turn by profiles of specific angular momentum flux in cylindrical radius ($\lambda$) by Reynolds stress (RS), meridional circulation (MC), viscous diffusion (VD), Maxwell stress (MS), and mean magnetic torques (MT), respectively. Terms are defined in detail in Equations~\ref{eq:AngMom1} and \ref{eq:AngMom2}. They are here averaged in time and longitude, and given in units of $10^{15}$ g s$^{-2}$. (b) Scalar plot of $z$-integrated fluxes of angular momentum with cylindrical radius $\lambda$ in units of $10^{38}$ g cm$^2$ s$^{-2}$ . Reynolds stress (RS, red) balance Maxwell stress (MS, blue), with viscous diffusion (VD, green) and magnetic torques (MT, purple) playing less of a role.  Contribution from the meridional circulation (MC, brown) are small. The sum of all five terms are also plotted (black dashed line).
  \label{fig:D3b amom}}
\end{center}
\end{figure*}

\section{Maintaining Rotational Shear}

A crucial component in the construction of magnetic wreaths is the strong latitudinal and radial shear from the differential rotation.  
The $\Omega$-effect has previously been shown to be the primary production mechanism for the magnetic wreaths in cases D3 and D5 \citep{Brown2010, Brown2011}, and it plays a key role in these simulations as well.  
Thus the angular momentum transport required to maintain the differential rotation is an important physical process in these dynamo models.
In the hydrodynamic models explored by \cite{Brown2008}, angular momentum transport in simulations at $3 \Omega_\odot$ was shown to be a balance between Reynolds stress supporting solar-like differential rotation with the meridional circulation and viscous diffusion tending to dissipate gradients in the rotation profile.  
With the addition of magnetic fields, Maxwell stress and mean magnetic torques can also transport angular momentum, changing the balance supporting the strong differential rotation achieved in the hydrodynamic cases.  
Even in cases without magnetic cycles such as case D3, \cite{Brown2010} showed that there are significant feedbacks on the differential rotation profile due to variations in the strength of the magnetic fields over time.
It is thus useful to examine not only the steady state balance of angular momentum transport over long time averages covering many magnetic cycles, but also to look at the temporal variability of those balances.

In order to explore the transport of angular momentum, let us examine the physical mechanisms which come into play. 
The balance of specific angular momentum along the rotation axis is determined by taking the product of the cylindrical radius $\lambda = r \sin{\theta}$ and the longitudinal component of the longitude-averaged momentum equation, which can be expressed as
 \begin{equation} 
 \frac{ \partial \mathcal{L}_z }{ \partial t } = \nabla \cdot \vec{\mathcal{F}} .
 \label{eq:Amom1}
 \end{equation}
We decompose the flux vector of mean angular momentum $\vec{\mathcal{F}}$ into radial and latitudinal components following prior convention \citep{Elliott2000, Brun2004, Brown2011}. 
We also decompose the flux vector into cylindrical coordinates along cylindrical radius ($\lambda$) and along the rotation axis ($z$), which in many ways is advantageous for displaying these quantities. 
A detailed description of this decomposition is given in Appendix B.   
The cylindrical flux of angular momentum is shown for case D3b over a long time average in Figure~\ref{fig:D3b amom}.  
The differential rotation is again clearly maintained by the Reynolds stress (RS), however here the terms opposing the differential rotation have changed compared to similar hydrodynamic cases.  
In case D3b the Maxwell stress (MS) is the largest term opposing the Reynolds stress with viscous diffusion (VD) and the mean magnetic torques (MT) each playing a small role, while contribution of the meridional circulation (MC) is almost insignificant.

 \begin{deluxetable}{lccccc}
  \tabletypesize{\footnotesize}
    \tablecolumns{6}
    \tablewidth{0pt}  
    \tablecaption{Production and Dissipation of Differential Rotation Kinetic Energy
    \label{table:DRKE}}
    \tablehead{\colhead{Case}  &  
      \colhead{$L_\mathrm{RS}$} &
      \colhead{$L_\mathrm{MC}$} &
       \colhead{$L_\mathrm{VD}$} &
      \colhead{$L_\mathrm{MS}$} &
      \colhead{$L_\mathrm{MT}$} 
   }
   \startdata
  D3     & 4.26 & -0.020 & -2.45 & -1.36 & -0.68 \\[3mm]
  D3a   & 3.18 & -0.032   & -1.11& -1.36 & -0.70 \\
  D3b   & 2.59 & -0.003 & -0.64 & -1.94 & -0.19 \\[3mm]
  D3-pm1 & 3.64 & -0.014 & -1.68 & -1.87 & -0.35 \\
  D3-pm2 & 3.32 & -0.024 & -1.61 & -1.95 & -0.31  \\
 \enddata
\tablecomments{Total production and dissipation of kinetic energy in the axisymmetric differential rotation profile over the entire simulated volume and averaged in time. Values for energy production rates are in units of $10^{32}$ erg s$^{-1}$. Production and dissipation terms are split following Equation~(\ref{eq:DRKE_prod}) into contributions from viscous dissipation, Reynolds stress, meridional circulations, Maxwell stress, and mean magnetic torques, respectively. Production terms are defined in Appendix B.
	}
\end{deluxetable}

  \begin{figure}
\begin{center}
 \includegraphics[width=0.9\linewidth]{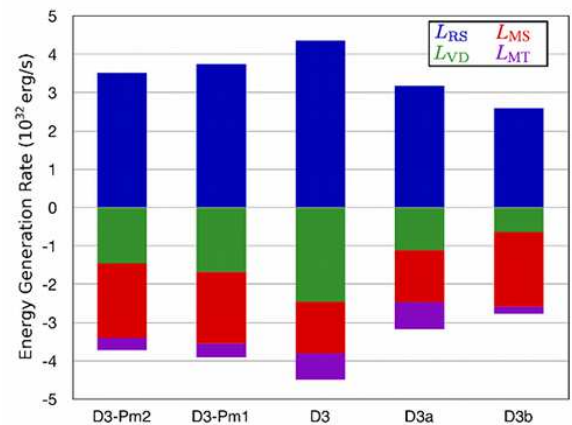}
 \caption{Companion to Table~\ref{table:DRKE}, showing the balance of time-averaged generation terms for the kinetic energy in the differential rotation profiles for each case indicated.  In all cases the differential rotation is maintained by a balance between the Reynolds stress and a combination of viscous diffusion and fluctuating and mean magnetic torques. The contribution from meridional circulations are not shown due to their small magnitude.   \label{fig:DRKE Balances}}
\end{center}
\end{figure}
 
We can write the evolution of the total energy of the differential rotation $ \mathcal{E}_\mathrm{DR} $ as
  \begin{equation}
\frac{\partial \mathcal{E}_{\mathrm{DR}} }{ \partial t }  = L_\mathrm{VD} + L_\mathrm{RS} + L_\mathrm{MC} + L_\mathrm{MS} + L_\mathrm{MT}  ,
\label{eq:DRKE_prod}
  \end{equation}
where the terms on the right-hand side represent the sources and sinks of kinetic energy in the differential rotation due to, respectively, viscous diffusion, Reynolds stress, meridional circulations, Maxwell stress, and mean magnetic torques.  
Appendix B provides a derivation of Equation~(\ref{eq:DRKE_prod}) and an expansion of the sources and sinks. 

Using this decomposition, we can examine the balance of production and dissipation of $\mathcal{E}_\mathrm{DR}$ averaged over long time intervals in each simulation.  
The balances are represented in Table~\ref{table:DRKE} and Figure \ref{fig:DRKE Balances}.  
For the increasing Pm branch the Reynolds stress change only slightly while the mean magnetic torques and viscous diffusion are systematically replaced by the Maxwell stress.  
Similar trends are observed in the constant Pm branch of cases, though here the magnitude of the Reynolds stress and viscous diffusion terms decrease more dramatically.  
This shift from unresolved dissipation in the form of SGS viscosity to resolved, small-scale torques from the Maxwell stress indicates that the balances which maintain strong differential rotation can persist  in less diffusive regimes, assuming that magnetic energies remain significantly smaller than kinetic energies.

Turning to the temporal variability in these balances, we find that for case D3b the departures from the values presented in Table~\ref{table:DRKE} and Figure~\ref{fig:DRKE Balances} are about 10\% for $L_\text{MS}$ and $L_\text{MT}$ when averaged over about 10 days, whereas those in $L_\text{RS}$, $L_\text{MC}$ and $L_\text{VD}$ are about 1\%.  This leads to decreases in the differential rotation when magnetic fields are strong, such as near the peak of the magnetic activity cycles.  Conversely, we observe modest increases in the differential rotation when magnetic fields are weak, such as during reversals of magnetic polarity.

\begin{figure}
\begin{center}
 \includegraphics[width=\linewidth]{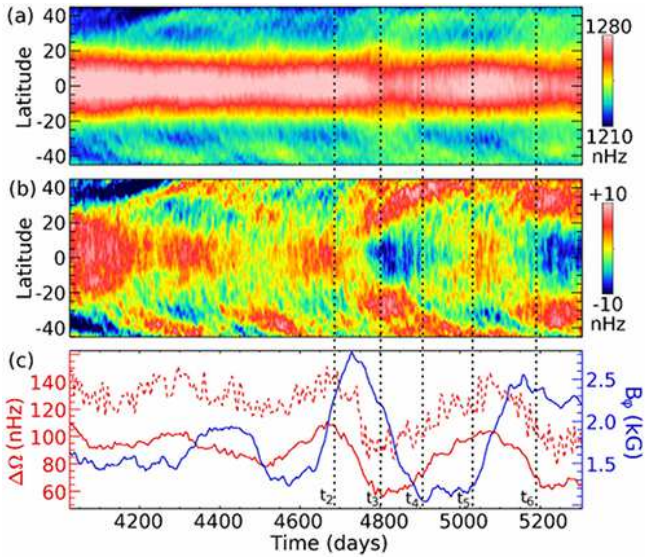}
 \caption{Temporal variability of differential rotation in case D3b over the same interval as in Figure~\ref{fig:Time Spec}. (a)~Longitudinally-averaged rotation rate at mid-convection zone as a function of time and latitude. (b)~Temporal variations are accentuated by subtracting the time-averaged $\Omega$ at each latitude.  Bands of faster rotating fluid move poleward on about the cycle period. (c)~Rotation contrasts in radius at the equator $\Delta_r$ (red, solid) and in latitude between the equator and $\pm 60^\circ$ in the upper convection zone $\Delta_\theta$ (red, dashed).  The volume-averaged toroidal field strength is also shown (blue, solid), with a phase-lag between peaks in magnetic field strength and decreases in differential rotation. Dotted lines indicate times $t_2$ through $t_6$ from Figure~\ref{fig:D3b reversal}.   \label{fig:DiffRotVar}}
\end{center}
\end{figure}

Figure~\ref{fig:DiffRotVar} shows the differential rotation in case D3b over several magnetic reversals. The differential rotation profile at mid-convection zone in Figure~\ref{fig:DiffRotVar}(a) is persistent, though there are small systematic variations in $\Omega$ revealed in Figure~\ref{fig:DiffRotVar}(b) during each magnetic cycle.  Figure~\ref{fig:DiffRotVar}(c) shows the differential rotation contrast in both radius and latitude over time as well as the volume-averaged toroidal magnetic field strength, indicating that the modest variations in the differential rotation are related to those in the magnetic field. 

\section{Production of magnetic field}

The transition from persistent wreaths in case D3 to cyclic wreaths and global polarity reversals in case D3b indicates that by reducing the levels of diffusion in these simulations we have fundamentally altered the balance of terms in the magnetic induction equation.  
The details of the reversal mechanism are likely to be very subtle in these highly nonlinear systems.  
In order to better understand the reversal mechanism, we explore the nature of the balances in the production and dissipation of toroidal and poloidal magnetic fields and provide indications of where and why changes in those balances are occurring, as well as some hints as to the nature of the reversal mechanism.

\subsection{Generation of Toroidal Magnetic Energy}

 In \cite{Brown2010} a detailed analysis of the balance of toroidal component of the axisymmetric induction equation was presented. 
 We write the toroidal component of the induction equation as 
\begin{equation}
\frac{ \partial B_\phi }{ \partial t }  = \left[ \nabla \times \left( \vec{v} \times \vec{B} \right) \right]_\phi - \nabla \times \left( \eta \nabla \times (B_\phi \hat{\phi}) \right) .
\label{eq:toroidal induction}
\end{equation}
Using vector identities, the first term on the right-hand side can be written as the sum of shearing terms, advection terms, and a compression term; additionally all of these terms can be decomposed into mean and fluctuating components \citep[for a full derivation, see Appendix A in][]{Brown2010}. 
That work also showed that the wreaths in case D3 are primarily generated by the $\Omega$-effect and dissipated by a combination of small-scale advection, shear, and diffusion.  

Here we perform a similar analysis, but instead of examining the generation of $\langle B_\phi \rangle$, we choose to examine the generation of the volume-integrated energy of the axisymmetric toroidal fields over the entire computational domain $\mathcal{V}$, given by
\begin{equation}
\mathcal{E}_\mathrm{TME} =  \int_{\mathcal{V}} \frac{ \langle B_\phi \rangle^2 }{ 8 \pi } \, d \mathcal{V} .
\end{equation}
We can construct an evolution equation for $\mathcal{E}_\mathrm{TME}$ by multiplying Equation~(\ref{eq:toroidal induction}) by $\langle B_\phi \rangle$.  
The result can be written as 
\begin{equation}
\frac{ \partial \mathcal{E}_\mathrm{TME} }{ \partial t} =  G_\mathrm{MS} + G_\mathrm{FS} + G_\mathrm{MA} + G_\mathrm{FA} + G_\mathrm{AC} + G_\mathrm{RD} ,
\label{eq:TME Production}
\end{equation}
where the six terms on the right-hand side represent, from left to right, the shearing of axisymmetric magnetic fields by mean flows associated with the $\Omega$-effect ($G_\mathrm{MS}$), the average of fluctuating flows shearing fluctuation fields ($G_\mathrm{FS}$), the advection of mean fields by mean flows ($G_\mathrm{MA}$), the average of fluctuating flows advecting fluctuating fields ($G_\mathrm{FA}$), the anelastic compression of fields ($G_\mathrm{AC}$), and the resistive diffusion of mean fields ($G_\mathrm{RD}$).  
Unlike in previous analyses which looked at the generation of magnetic field vectors, here we are concerned with scalar quantities. 
The terms on the right-hand side of Equation~(\ref{eq:TME Production}) are computed as 
\begin{equation}
G_\mathrm{MS} =  \int_{\mathcal{V}}  \langle B_\phi \rangle \left[ \left( \langle \vec{B} \rangle \cdot \nabla \right) \langle \vec{v} \rangle \right]_\phi \, d \mathcal{V} ,
\label{eq:omega effect}
\end{equation}
\begin{equation}
G_\mathrm{FS} =  \int_{\mathcal{V}} \langle B_\phi \rangle \left[ \Big\langle \left( \vec{B}^\prime \cdot \nabla \right) \vec{v}^\prime \Big\rangle \right]_\phi \, d \mathcal{V} ,
\end{equation}
\begin{equation}
G_\mathrm{MA} = - \int_{\mathcal{V}} \langle B_\phi \rangle \left[ \left( \langle \vec{v} \rangle \cdot \nabla \right) \langle \vec{B} \rangle \right]_\phi \, d \mathcal{V} ,
\end{equation}
\begin{equation}
G_\mathrm{FA} = - \int_{\mathcal{V}} \langle B_\phi \rangle \left[ \Big\langle \left( \vec{v}^\prime \cdot \nabla \right) \vec{B}^\prime \Big\rangle \right]_\phi \, d \mathcal{V} ,
\end{equation}
\begin{equation}
G_\mathrm{AC} =  \int_{\mathcal{V}} \langle B_\phi \rangle \Big\langle v_r B_\phi \frac{ \partial \ln{\bar{\rho}} }{ \partial r } \Big\rangle \, d \mathcal{V} ,
\end{equation}
\begin{equation}
G_\mathrm{RD} = - \int_{\mathcal{V}} \langle B_\phi \rangle \nabla \times \left( \eta \nabla \times \langle B_\phi \rangle \right) \, d \mathcal{V} .
\end{equation}
For consistency, angle brackets denote longitude averages.

\begin{figure}
\begin{center}
 \includegraphics[width=0.95\linewidth]{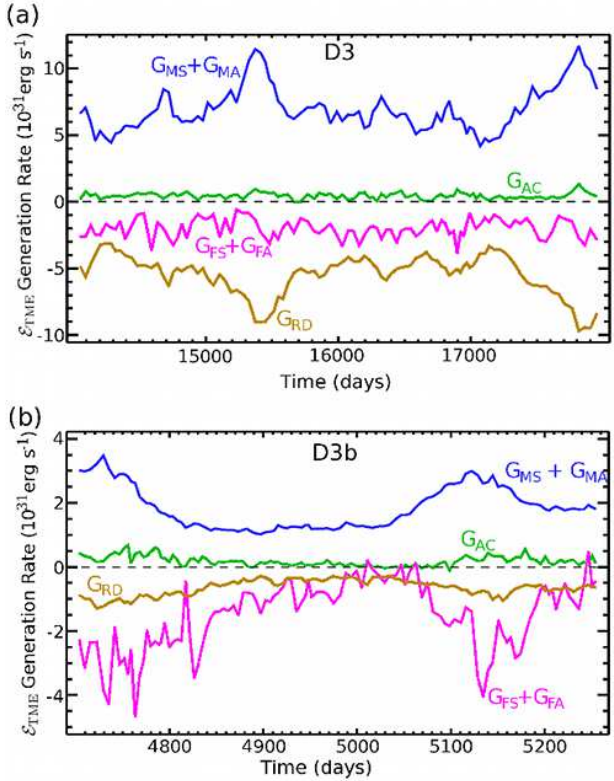}
 \caption{Volume-integrated production terms of magnetic energy in the mean toroidal fields from Equation~(\ref{eq:TME Production}) for (a) case D3 and (b) case D3b. We have combined the mean shear and mean advection terms (blue line) and the fluctuating shear and fluctuating advection (purple line). In both simulations, energy is produced primarily by the shearing of mean fields by mean flows.  Also in both cases compression of fields (green line) plays a very small role.  In case D3 diffusion (red line) and the advection and shear of fluctuating flows on fluctuating fields destroy energy, with diffusion generally a factor of 2.5 larger.  In case D3b, however, the dissipation of energy by fluctuating advection and shear is 2.2 times greater on average than diffusion.  Thus in case D3b resolved turbulence is the primary mechanism for dissipating the magnetic wreaths.
\label{fig:Mag Energy Production}}
\end{center}
\end{figure}

Figure~\ref{fig:Mag Energy Production} shows the temporal evolution of the integrated energy of the axisymmetric toroidal field for cases D3 and D3b and the behavior of the production terms governing the variation of $\mathcal{E}_\mathrm{TME}$.  
We have chosen to combine the contributions of the mean shear and advection terms and the fluctuating shear and advection terms for ease of viewing.  
The mean advection term $G_\mathrm{MA}$ is generally positive and always much smaller than the mean shear term $G_\mathrm{MS}$.  
The fluctuating shear and advection terms are both generally negative, of approximately the same magnitude, and tend to vary in phase with each other.

Let us first look at the average levels of each term plotted in Figure~\ref{fig:Mag Energy Production} to get a sense for the basic balance of terms. 
The production of $\mathcal{E}_\mathrm{TME}$ is dominated by the mean shear term which is large and always positive in both case D3 and D3b.  
The compression term in both cases is roughly an order of magnitude smaller but is again always positive due to the asymmetry in upflows and downflows in compressible convection, which gives preference to downward pumping of magnetic field causing an increase in magnetic energy due to compression.  
The production of magnetic energy is opposed by the resistive diffusion, fluctuating advection, and fluctuating shear terms. 
In case D3 resistive diffusion is roughly three times larger than the sum of the two fluctuating terms, while in case D3b the roles are reversed and resolved turbulent dissipation does most of the destruction of $\mathcal{E}_\mathrm{TME}$ while the unresolved turbulent dissipation represented by our explicit resistivity is relegated to a less prominent role.  
Supporting this transition from unresolved to resolved dissipative processes, in case D3 the sum of the fluctuating terms does not show noticeable changes in behavior when the magnetic energy is high versus when it is low.  
Instead the response is seen primarily in the resistive dissipation term.  
In case D3b, however, the fluctuating terms show strong variations in response to changes in the magnetic energy.

\begin{figure*}
\begin{center}
 \includegraphics[width=0.9\linewidth]{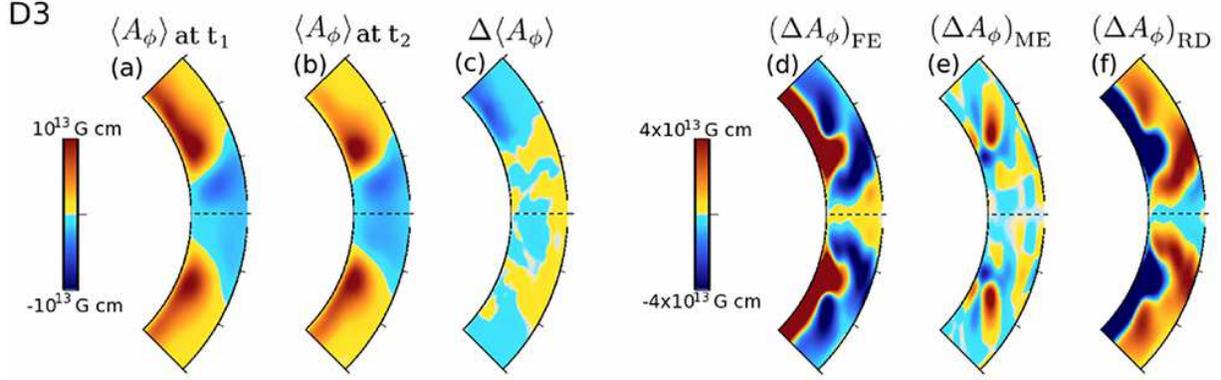}
 \caption{Time-evolution of $\langle A_\phi \rangle $ between $\pm 45^\circ$ latitude for case D3 over about 500 days.  Times $t_1$ and $t_2$ for case D3 correspond to times indicated in Figure~\ref{fig:TimeLat Stack}.  Shown are $ \langle A_\phi \rangle$ at (a) the beginning and (b) end of the time interval, (c) the net change between those times $\Delta \langle A_\phi \rangle$, the changes in $ \langle A_\phi \rangle$ due to (d) the fluctuating EMF $\left( \Delta A_\phi \right)_{\mathrm{FE}} $, (e) the mean EMF $\left( \Delta A_\phi \right)_{\mathrm{ME}}$, and (f) resistive diffusion $\left( \Delta A_\phi \right)_{\mathrm{RD}}$.  Of particular importance is the region of positive production in $\left( \Delta A_\phi \right)_{\mathrm{FE}}$, which if left unimpeded by diffusion would lead to a reversal in global magnetic polarity. The color table has been chosen with a sharp transition from light blue to yellow around zero, thus low-amplitude signals, such as seen in (c) and (e), are highlighted.
 \label{fig:Aphi Balance D3}}
\end{center}
\end{figure*}

\begin{figure*}
\begin{center}
 \includegraphics[width=0.9\linewidth]{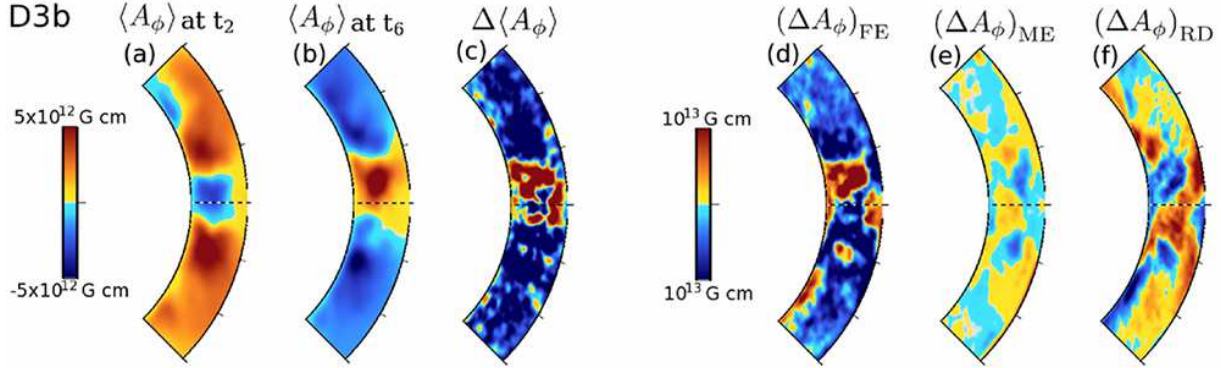}
 \caption{Same Figure~\ref{fig:Aphi Balance D3}, but for case D3b.  Times $t_2$ and $t_6$ for case D3b correspond to times indicated in Figures~\ref{fig:TimeLat Stack} and \ref{fig:Time Spec}. The turbulent EMF induces field of the opposite sense to that which was present at $t_2$ and is opposed by the resistive diffusion.  Note that $ \left( \Delta A_\phi \right)_{\mathrm{FE}}$ and $ \left( \Delta A_\phi \right)_{\mathrm{RD}}$ for both cases are topologically similar, but that $ \left( \Delta A_\phi \right)_{\mathrm{RD}}$ is smaller in case D3b, rendering it unable to prevent the reversal of $\langle A_\phi \rangle$ by the fluctuating EMF which begins with the positive region near the equator.
\label{fig:Aphi Balance D3b}}
\end{center}
\end{figure*}

This transition from wreath-building dynamos that rely on our SGS diffusion to wreath-building dynamos that are sufficiently turbulent to be dominated by resolved turbulent dissipation answers one important question relative to the extension of this dynamo mechanism to even more turbulent states.  
It had long been postulated that global-scale magnetic structures could not exist in the convection zone as they would be quickly destroyed by the intense turbulence.  
While it is clearly possible that our wreaths may not be able to survive if we were able to simulate far more turbulent conditions, case D3b marks an important milestone along the path towards the possibility of magnetic wreaths coexisting with highly turbulent convection.

Returning to Figure~\ref{fig:Mag Energy Production}, let us now look at the time-variation in the production terms.  
Both cases show variability of $\mathcal{E}_\mathrm{TME}$, but for case D3b we have chosen to show a time period that includes states before, during, and after a reversal in global magnetic polarity.  
For both cases, careful examination shows that changes in $\mathcal{E}_\mathrm{TME}$ are initiated primarily by changes in $G_\mathrm{MS}$, not by changes to the terms dissipating energy.  
The terms representing both resolved and unresolved diffusion respond to changes in $\mathcal{E}_\mathrm{TME}$ rather than drive them.  
In case D3 this is supported by the cross-correlation of $G_\mathrm{MS}$ and $G_\mathrm{RD}$ peaking at a 39 day lag, while there is no significant cross-correlation between $G_\mathrm{MS}$ and either $G_\mathrm{FS}$ or $G_\mathrm{FA}$ for any shift in time.  
In case D3b both the cross-correlation of $G_\mathrm{MS}$ with $G_\mathrm{FS}$, and $G_\mathrm{MS}$ with $G_\mathrm{FA}$ both peak at a lag of 11 days.  
Resistive diffusion responds faster in case D3b with a peak in cross-correlation for a lag of only 5.6 days.  
This demonstrates that the variability in the toroidal fields is driven by changes in the generation of field by the $\Omega$-effect.  

If we more closely examine the structure of $G_\mathrm{MS}$ from Equation~(\ref{eq:omega effect}), we can expand it to
\begin{align}
G_\mathrm{MS} = \int_{\mathcal{V}} & \left( \langle B_\phi \rangle \langle B_r \rangle \frac{ \partial \langle v_\phi \rangle }{ \partial r } + \frac{ \langle B_\phi \rangle \langle B_\theta \rangle }{ r } \frac{ \partial \langle v_\phi \rangle }{ \partial \theta } \right. \nonumber \\
& + \left. \frac{ \langle B_\phi \rangle^2 \langle v_r \rangle }{ r } + \frac{ \langle B_\phi \rangle^2 \langle v_\theta \rangle }{ r \; \tan{\theta} } \right)  \, d \mathcal{V} .
\label{eq:omega effect 2}
\end{align}
The third and fourth terms are geometric terms from the spherical coordinate system which are generally small.  
In order to produce a change in $G_\mathrm{MS}$, the dynamo can either change the axisymmetric poloidal field or modify the differential rotation of the domain.  
We have examined both the amplitude and geometry of the mean shear due to differential rotation and find only very small changes in any of the cases presented here.  
Additionally, reversals in the polarity of the wreaths such as those seen in cases D3a and D3b require a change in sign for the generation term (obtained by dividing by $\langle B_\phi \rangle$) and there is never a change of sign in the shear profile of the differential rotation observed in any of these cases.  
Thus we are left with the conclusion that reversals in the polarity of the axisymmetric toroidal fields must be initiated by changes in the axisymmetric poloidal fields.

\subsection{Collapse of Resistive Balance Leading to Reversals}

The key to understanding the reversals seen in cases D3a and D3b lies in the generation of poloidal field.  
When the poloidal field reverses sign the $\Omega$-effect can then build wreaths of the opposite polarity and reverse the sign of the axisymmetric toroidal field.  
It is difficult to identify a simple model for the generation of poloidal field in these cases, particularly in case D3b.  
We can, however, identify the change in the generation mechanism that occurred between cases D3 and D3b.

Following the work of \cite{Brown2010, Brown2011}, we choose to examine the evolution of the $\phi$ component of the mean vector magnetic potential $\langle \vec{A} \rangle$.  
This is convenient as $\langle A_\phi \rangle$ completely specifies the components of the axisymmetric poloidal magnetic field by
  \begin{equation}
 \nabla \times \left( \langle A_\phi \rangle \hat{\phi} \right)  = \langle B_r \rangle \hat{r} + \langle B_\theta \rangle \hat{\theta} .
 \label{eq:define Aphi}
  \end{equation} 
The temporal variations in the magnetic wreaths are driven by changes in the shear of mean poloidal magnetic fields by mean differential rotation and that only the axisymmetric poloidal fields can change sign, hence changes in the polarity of the wreaths can be traced back to the evolution of $\langle A_\phi \rangle$.  
Further, the key region of the domain in which we should monitor $\langle A_\phi \rangle$ is near the equator where the gradients in differential rotation are largest and where the wreaths are primarily generated.

The evolution of $\langle A_\phi \rangle$ is governed by 
  \begin{equation}
  \frac{ \partial \langle A_\phi \rangle }{ \partial t } = \left( \langle \vec{v} \rangle \times \langle \vec{B} \rangle \right)_\phi + \left( \langle v'  \times B'  \rangle \right)_\phi - \eta \langle J_\phi \rangle .
  \label{eq:evolve Aphi}
  \end{equation} 
We have ignored a gauge term in Equation (\ref{eq:evolve Aphi}) which is permissible for any longitudinally-periodic gauge. 
We take a time-integral of this equation to look at the changes in $\langle A_\phi \rangle$ over about 500 days in cases D3 and D3b, and define the time-integral of each term as 
  \begin{equation}
 \left( \Delta A_\phi \right)_{\mathrm{ME}} =  \int^{t_2}_{t_1} \left( \langle \vec{v} \rangle \times \langle \vec{B} \rangle \right)_\phi dt
 \end{equation}
 \begin{equation}
  \left( \Delta A_\phi \right)_{\mathrm{FE}} =  \int^{t_2}_{t_1} \left( \langle v'  \times B'  \rangle \right)_\phi dt
\end{equation}
\begin{equation}
 \left( \Delta A_\phi \right)_{\mathrm{RD}} = - \int^{t_2}_{t_1}  \eta \langle J_\phi \rangle dt .
  \end{equation} 
Thus the change in $ \langle A_\phi \rangle$ can be written as 
\begin{equation}
\Delta \langle A_\phi \rangle = \left( \Delta A_\phi \right)_{\mathrm{FE}} + \left( \Delta A_\phi \right)_{\mathrm{ME}} + \left( \Delta A_\phi \right)_{\mathrm{RD}} .
  \end{equation} 

Figures~\ref{fig:Aphi Balance D3} and \ref{fig:Aphi Balance D3b} show the evolution of $\langle A_\phi \rangle$ in cases D3 and D3b, respectively, as well as the time-integrated production of terms shown above and the net change over the time interval. 
For case D3b we chose a time period spanning a reversal in global magnetic polarity.  
In both cases $\left( \Delta A_\phi \right)_{\mathrm{ME}}$ is small and the evolution is primarily governed by the balance between fluctuating EMF and resistive diffusion. 
The primary difference between cases D3 and D3b is the collapse of the resistive balance.  
Both cases show similar patterns in $\left( \Delta A_\phi \right)_{\mathrm{FE}}$, namely that the fluctuating EMF in both cases is seeking to create a region of opposite polarity poloidal field near the equator while reinforcing the current sense of poloidal field at mid-latitudes.  
Thus in both cases D3 and D3b the turbulent correlations between the existing field and the convective turbulence tends to build poloidal field near the equator of the opposite sense than the field that built the current wreaths through the $\Omega$-effect. 
The difference between cases D3 and D3b is that in case D3 the diffusion term is sufficiently large to prevent the reversal by diffusing away the opposite polarity poloidal field at the equator before it can accumulate sufficiently to cause a reversal.

  \begin{figure}
\begin{center}
 \includegraphics[width=0.75\linewidth]{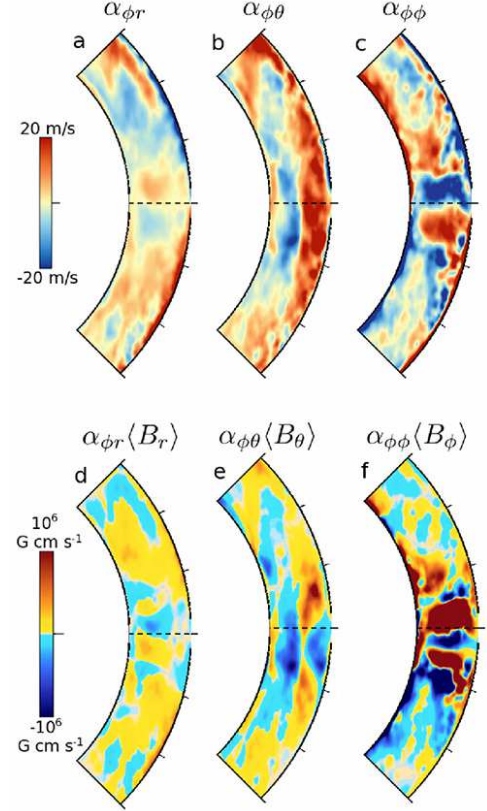}
 \caption{(a-c)~Values for the three components of $\alpha$ tensor relevant to the generation of $\langle \epsilon^{\prime}_\phi \rangle$ as a function of radius and latitude.  Values are computed using a singular value decomposition over approximately 3000 days of simulation time with the assumption that these components of $\alpha$ are spatially local and do not vary in time. The ${\bf \alpha}_{\phi r}$ component is very small, whereas the $\alpha_{\phi \theta}$ and $\alpha_{\phi \phi}$ components show significant spatial variability and comparable amplitude. (d-f)~Values for components of $\alpha_{\phi j} \langle B_j \rangle$, showing the effect of each component on $\langle \epsilon^{\prime}_\phi \rangle$. Magnetic fields have been averaged over the same interval as in Figure~\ref{fig:Aphi Balance D3b} (about 480 days). Here the contribution of $\alpha_{\phi \phi} \langle B_\phi \rangle$ is dominant, with a smaller but still significant contribution by $\alpha_{\phi \theta} \langle B_\theta \rangle$. 
\label{fig:alpha Tensor}}
\end{center}
\end{figure}

What causes the fluctuating EMF to display this propensity towards reversing the polarity of $\langle A_\phi \rangle$ near the equator? 
It would seem that there should be some link back to the strong toroidal wreaths, however when we expand the fluctuating EMF, we find that
  \begin{equation}
\left( \langle v'  \times B'  \rangle \right)_\phi = \langle v^\prime_r B^\prime_\theta - v^\prime_\theta B^\prime_r \rangle .
\label{eq:poloidal correlations}
  \end{equation} 
  Clearly, neither the axisymmetric nor fluctuating components of $B_\phi$ come into play here, indicating that to complete a reversal we need to connect the large-scale toroidal fields to correlations between small-scale poloidal fields and poloidal flows.  
As shown in Figure~\ref{fig:meet D3 family}(i), the wreaths are not purely toroidal structures, thus the small-scale fields needed in Equation~\ref{eq:poloidal correlations} may be supplied by the wreaths themselves.
However, we have not been able to definitively link the poloidal components of the wreaths to the reversal process.
While the subtle nature of this process remains difficult to pin down, we do have some hints at its origin.

\subsection{Exploring An $\alpha$-Like Effect}

The final step in the reversal process is what is often described in the parlance of mean-field dynamo theory as the $\alpha$-effect \citep[see][]{Charbonneau2010}.  
Generally, the $\alpha$-effect is the source of the axisymmetric component of the turbulent EMF, defined as 
  \begin{equation}
\langle \vec{\epsilon'} \rangle = \langle \vec{v'}  \times \vec{B'}  \rangle.
  \end{equation} 
Specifically, we are interested in the zonal component which generates the mean poloidal field and its connection to the axisymmetric toroidal field, which might be expressed as
\begin{equation}
\langle \epsilon^{\prime}_\phi \rangle = \alpha_{\phi r} \langle B_r \rangle + \alpha_{\phi \theta} \langle B_\theta \rangle + \alpha_{\phi \phi} \langle B_\phi \rangle .
\label{eq:tensor alpha}
\end{equation}
In its simplest formulation, the components of $\alpha_{i j}$ in Equation~(\ref{eq:tensor alpha}) are constants, however more complex formulations exist.

For case D3b, we have computed the value of the three components of the $\alpha$ tensor in Equation~(\ref{eq:tensor alpha}) using a singular value decomposition following the work of \cite{Racine2011}. 
We compute values for $\alpha_{ij}$ at each radial and latitudinal location, assuming that $\alpha_{ij}$ is constant in time.
The results of Figure~\ref{fig:alpha Tensor} demonstrate that the $\alpha_{\phi \phi} \langle B_\phi \rangle$ is the most important term in the generation of the fluctuating toroidal EMF.
Thus, it is particularly intriguing to focus on the connection 
\begin{equation}
\langle \epsilon^{\prime}_\phi \rangle = \alpha_{\phi \phi} \langle B_\phi \rangle .
\label{eq:alpha}
\end{equation}
\cite{Brown2010} showed that for one formulation of an $\alpha$-effect in case D3, $\alpha_{\phi \phi}$ was spatially nonlocal, which would not be picked up in our fitting procedure. 

  \begin{figure}
\begin{center}
 \includegraphics[width=0.9\linewidth]{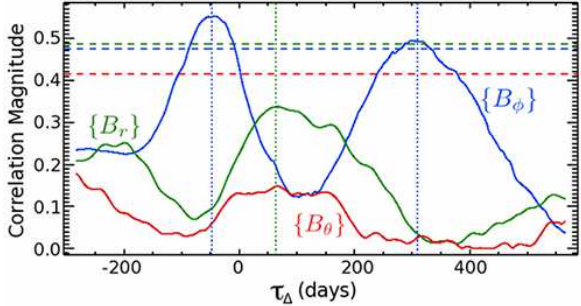}
 \caption{Magnitude of cross-correlation in time of $\{ \epsilon^{\prime}_\phi \}$ and $\{ B_r \}$ (green), $\{ B_\theta \}$ (red), and $\{ B_\phi \}$ (blue) for case D3b.
 Cross-correlation is computed as a function of the temporal offset $\tau_{\Delta}$,/ with negative offsets indicating magnetic fields precede the toroidal EMF. Also shown are the $2 \sigma$ confidence levels (dashed), computed using a Markov chain Monte Carlo method \citep{Wall2003}. The only statistically significant peaks are those relating $\{ \epsilon^{\prime}_\phi \}$ and $\{ B_\phi \}$. 
\label{fig:XCorrel}}
\end{center}
\end{figure}

The exact mechanism for connecting mean fields and the fluctuating EMF is subtle, but we find that in case D3b an $\alpha$-like effect emerges, which is nonlocal in time, acting on the same time scale as convective overturning. 
If we consider correlations between the volume-averaged magnetic field components and similarity the fluctuating toroidal EMF, we find evidence that the $\alpha$-like effect in case D3b is not instantaneous but rather acts on a time scale (47 days) which is commensurate with the convective overturning time.  
The volume averages, denoted by curly braces, are computed separately for each hemisphere over all depths and longitudes, and between the equator and $\pm 30^\circ$ in latitude.  
Combining the data for both hemispheres, the cross-correlation is computed and shown in Figure~\ref{fig:XCorrel} as a function of the temporal interval $\Delta_\tau$ by which $\{ \epsilon^{\prime}_\phi \}$ is offset relative to $\{ B_r \}$, $\{ B_\theta \}$, and $\{ B_\phi \}$ in turn.
The peaks in the cross-correlation which exceed $2 \sigma$ in significance occur when $\{ \epsilon^{\prime}_\phi \}$ leads $\{ B_\phi \}$ by 312 days and when $\{ B_\phi \}$ leads $\{ \epsilon^{\prime}_\phi \}$ by 47 days.

  \begin{figure}
\begin{center}
 \includegraphics[width=\linewidth]{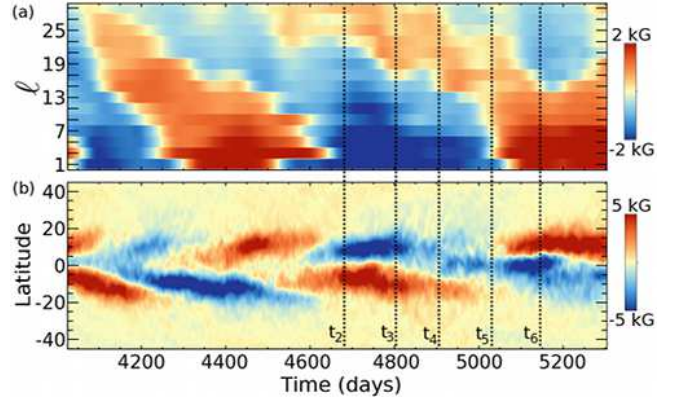}
 \caption{Companion plots of the time evolution in case D3b of (a) the spherical harmonic coefficients for antisymmetric modes with $1 \le \ell \le 29$ and $m = 0$ for $B_\phi$, and (b) $\langle B_\phi \rangle$ in physical space as a function of latitude, both at mid-convection zone. Dashed lines show times referenced in Figures~\ref{fig:TimeLat Stack}, \ref{fig:D3b reversal}, \ref{fig:Aphi Balance D3}, and \ref{fig:Aphi Balance D3b}. A factor of  $\left( -1 \right)^{(\ell-1)/2}$ is applied to the spherical harmonic coefficients to remove the effect of the wreaths confinement to low latitudes. There is a clear progressive spectral transfer of magnetic energy from high $\ell$ modes to low $\ell$ modes as each cycle progresses.  Reversals begin at moderate scales (high $\ell$) and then progress to large scales (low $\ell$).
\label{fig:Time Spec}}
\end{center}
\end{figure}

Analysis of the autocorrelation of both $\{ B_\phi \}$ and$\{ \epsilon^{\prime}_\phi \}$ indicates that the two peaks are not due to periodicities in either of the two time series individually.  
Further, the widths of these peaks largely originates from the coherence time for $\langle B_\phi \rangle$ of about 100 days.
The first peak at 312 days represents the time scale for the $\Omega$-effect and agrees well with the estimate from mean-field theory $\tau_\Omega$ given by
  \begin{equation}
 \tau_\Omega = P_\text{rot} \frac{\Omega_0}{\Delta \Omega} \frac{\langle B_\phi \rangle}{\langle B_\text{pol} \rangle} , 
  \end{equation} 
where $P_\text{rot}$ is the rotation period, $\Omega_0$ is the frame rotation rate, $\Delta \Omega$ is the differential rotation rate, and $\langle B_\text{pol} \rangle$ is the strength of the poloidal field. For case D3b, this yields a value of 324 days. 
The second peak in the cross-correlation between the two fields occurs when $\tau_\Delta = -47$ days. 
This peak suggests that the correlations which generate the turbulent zonal EMF are related in some way to the axisymmetric toroidal fields, and that whatever mechanism establishes this correlation, it has a time scale of about 50 days. 
This temporally and spatially nonlocal $\alpha$-effect clearly points to convection as a key player, as other mechanisms like the meridional circulation are at least an order of magnitude slower.

\begin{figure*}
\begin{center}
 \includegraphics[width=0.7\linewidth]{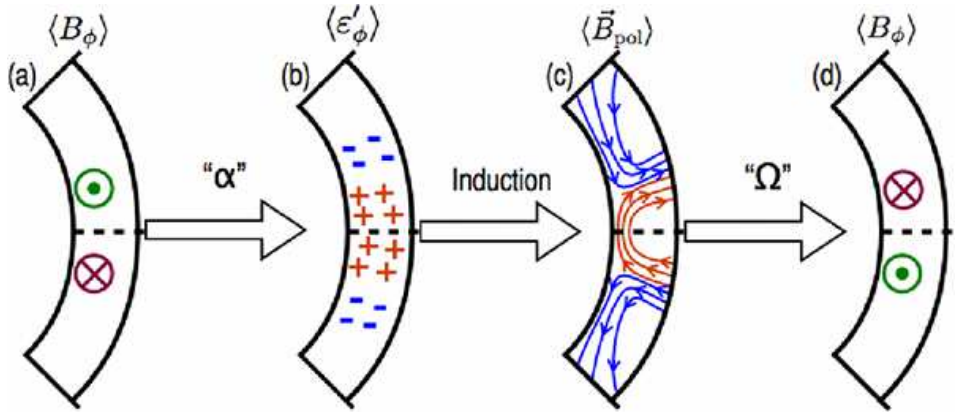}
 \caption{ Schematic description of the reversal mechanism for cyclic convective dynamos in four steps.  (a) Two toroidal wreaths at low latitude which generates a turbulent EMF via a nonlocal ``$\alpha$"-effect, either through nonlinear interactions across the equator or via helical convection.  The sign of the EMF changes at roughly the location of the wreaths.  (b) Correlations in turbulent poloidal velocities and fluctuating magnetic field drive an induction of mean poloidal field which is roughly octopolar. (c) Mean poloidal field near the equator is sheared by differential rotation to generate mean toroidal field through the $\Omega$-effect.  In these simulations, the largest component is the shearing of radial field lines by radial gradients in the differential rotation. (d) Toroidal wreaths of opposite polarity are generated.
\label{fig:Reversal Cartoon}}
\end{center}
\end{figure*}

In addition to a timescale for an $\alpha$-like effect which is commensurate with the convective overturning time, we also find evidence for an upscale transfer of magnetic energy related to magnetic reversals. 
Figure~\ref{fig:Time Spec} shows the temporal evolution of $\langle B_\phi \rangle$ at mid-convection zone over both latitude and spherical harmonic degree $\ell$.  
Several reversals of global magnetic polarity are evident in Figure~\ref{fig:Time Spec}(b), including the reversal shown in detail in Figure~\ref{fig:D3b reversal}. 
Figure~\ref{fig:Time Spec}(a) shows the coefficients of the spherical harmonic expansion of the axisymmetric toroidal field for the anti-symmetric (odd $\ell$) modes with $1 \le \ell \le 29$ over roughly three magnetic activity cycles. 
In both physical space and spectral space, it is clear that each cycle has opposite polarity from the preceding cycle.

There is a  preference for antisymmetric modes with odd values of $\ell$, as would be expected from the $\Omega$-effect acting on a poloidal field that is preferentially symmetric about the equator (even $\ell$).  
The upscale cascade involving odd modes is expected from both theoretical and observational studies families of dynamo modes \citep[see][]{Nishikawa2008, DeRosa2012}. 
As a reversal occurs we see power showing up first at moderate $\ell$ and then cascading upscale to smaller $\ell$ values until it peaks at $\ell = 3$ or 1 depending on the cycle. 
The reversal starts at $25 \lesssim \ell \lesssim 29$ and then each successive mode reverses. 
There is considerable overlap between cycles, in some cases reversals are seen in the high-$\ell$ modes in as little as a hundred days after the previous reversal is completed at low-$\ell$. 
We note that convective power peaks at spherical harmonic degrees between about 25 and 40 in these simulations.  
This suggests that the reversals are caused by turbulent processes interacting with the wreaths, yielding an upscale energy transfer which organizes the large-scale fields.
Combined with our cross-correlation analysis, this upscale transfer indicates the key role of convection in connecting mean toroidal magnetic fields with the fluctuating toroidal EMF.

As illustrated schematically in Figure~\ref{fig:Reversal Cartoon}, the reversal mechanism involves three main processes. 
First, axisymmetric wreaths of toroidal magnetic field (Figure~\ref{fig:Reversal Cartoon}(a)) lead to correlations in the non-axisymmetric poloidal velocity and magnetic fields which drive an axisymmetric turbulent EMF through an $\alpha$-like effect. 
The upscale transfer of magnetic energy and the fact that the correlation between the magnetic energy of the wreaths and the turbulent EMF peaks on roughly a convective overturning time would seem to point towards the convective motions as a key player in this $\alpha$-like process. 
In the second step, the turbulent EMF reinforces the dominant poloidal field at mid-latitudes but is the opposite sign near the equator (Figure~\ref{fig:Reversal Cartoon}(b)), creating an octopolar configuration, with strong radial  field concentrations at low latitudes (Figure~\ref{fig:Reversal Cartoon}(c)).  
As the reversal progresses, the region of new poloidal field shown in red in Figure~\ref{fig:Reversal Cartoon}(c) will expand and eventually replace the old sense of field shown in blue.
The third step involves axisymmetric poloidal magnetic field being sheared by differential rotation.  
Here the differential rotation is largely cylindrical, thus radial poloidal field is primarily converted into toroidal magnetic field through the $\Omega$-effect, which results in axisymmetric toroidal fields of the opposite polarity (Figure~\ref{fig:Reversal Cartoon}(d)). 
The process then repeats with the opposite polarity.

\section{Turbulence-Regulated Flux Emergence}

Photospheric active regions are thought to arise from the buoyant destabilization, rise, and emergence of coherent, subsurface toroidal flux structures.  
It is often argued that these subsurface flux structures originate below the convection zone, where the strong shear of the tachocline promotes toroidal flux generation and the subadiabatic stratification of the overshoot region promotes flux storage by inhibiting the buoyancy instability \citep{Galloway1981, VanBallegooijen1982}.
In this section we offer an alternative viewpoint that is inspired and supported by the numerical models presented here.  
Namely, we argue that buoyant flux structures may be produced in the Sun and stars not only in the tachocline but also in the lower convection zone through the combined action of rotational shear and turbulent intermittency.  

In previous papers we have demonstrated that organized systems of toroidal flux can persist within a turbulent convection zone despite the inhibiting influence of turbulent dispersal \citep{Brown2010, Brown2011}.  
Here we have demonstrated that this continues to hold as we decrease the diffusion, crossing a threshold beyond which resolved motions replace artificial dissipation in the dynamical balances that sustain mean flows and fields.  
Furthermore, as the diffusion is decreased, intense, localized wreath cores form where the magnetic energy density exceeds the surrounding kinetic energy density (Figure~\ref{fig:Bphi PDFs}).  
This trend is highlighted most dramatically by case S3, where the much lower diffusion promotes coherent wreath cores strong enough to become buoyant, as first demonstrated by \cite{Nelson2011}.

Here we explore in more general terms the link between magnetic wreaths and flux emergence, addressing in particular on how it might operate in real stars where the dissipation is many orders of magnitude less than in simulations.  
We begin by noting that the $\Omega$-effect does not just operate on axisymmetric fields; poloidal fields of all longitudinal wavenumbers ($m$) in the convection zone are converted to toroidal fields (of the same $m$) and amplified by rotational  shear, blurring the distinction between mean and fluctuating fields.
Turbulent intermittency in the surrounding convection can further amplify shear-generated flux structures, promoting the generation of fibril magnetic fields and coherent, localized wreath cores (Figure~\ref{fig:Bphi PDFs}).

The low Mach number of stellar convection zones ensures that the gas pressure adjusts rapidly to any imbalance of mechanical and magnetic stresses.  
Thus, the formation of fibril, intermittent flux concentrations (wreath cores) will induce a pressure perturbation $\delta P \sim P_t - P_m$, where $P_t$ is the turbulent (kinetic plus magnetic) pressure of the surrounding medium and $P_m$ is the magnetic pressure associated with the coherent flux that defines the wreath core.  
We have neglected the turbulent pressure within the wreath core which may be suppressed by magnetic tension, providing a positive feedback mechanism that can further promote the formation of coherent, superequiparition wreath cores and buoyant loops \citep{Kleeorin1989, Rogachevskii2007, Kapyla2012}.

Weak magnetic flux concentrations, $P_m < P_t$, are not susceptible to buoyancy instabilities because their magnetic pressure is insufficient to balance the surrounding turbulent pressure, resulting in $\delta P > 0$.  
It is only the strongest wreath cores that develop a pressure deficit $\delta P < 0$, in particular only those cores in which the magnetic pressure  $P_m$ exceeds the stabilizing influence of the surrounding convective motions.  
This implies that a necessary but not sufficient condition for the wreath cores to become buoyant is that they must be superequipartition relative to the surrounding  convection.  
The surrounding flows may in turn enhance or retard the tendency for such structures to rise.
As demonstrated in Figure~\ref{fig:Bphi PDFs}, this is indeed achieved in our simulations and it becomes more pronounced as the artificial diffusion is reduced, eventually inducing buoyant rise.  

\begin{figure}
\begin{center}
 \includegraphics[width=0.85\linewidth]{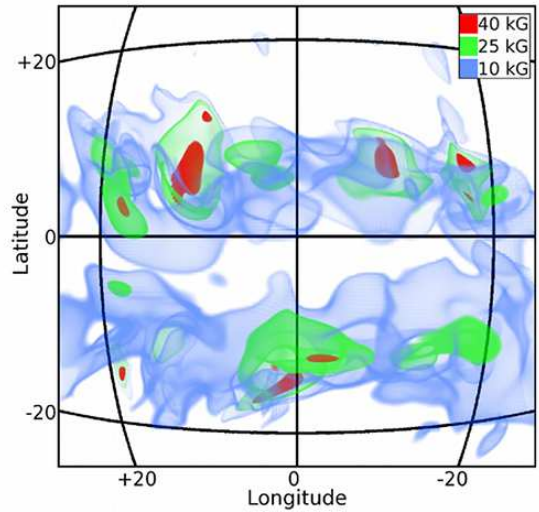}
 \caption{Three-dimensional volume renderings of isosurfaces of magnetic field amplitude in case S3.  Blue surfaces have amplitudes of 10 kG, green surfaces represent 25 kG, and red surfaces indicate 40 kG fields.  Grid lines indicate latitude and longitude at 0.72 $R_\odot$ as they would appear from the vantage point of the viewer.  Small portions of the cores of these wreaths have been amplified to field strengths in excess of 40 kG while the majority of the wreaths exhibit fields of about 10 kG or roughly in equipartition with the mean kinetic energy density (see Figure~\ref{fig:Bphi PDFs}).
 \label{fig:S3 Iso}
 \vspace{-0.5cm}}
\end{center}
\end{figure}

\begin{figure*}
\begin{center}
 \includegraphics[width=\linewidth]{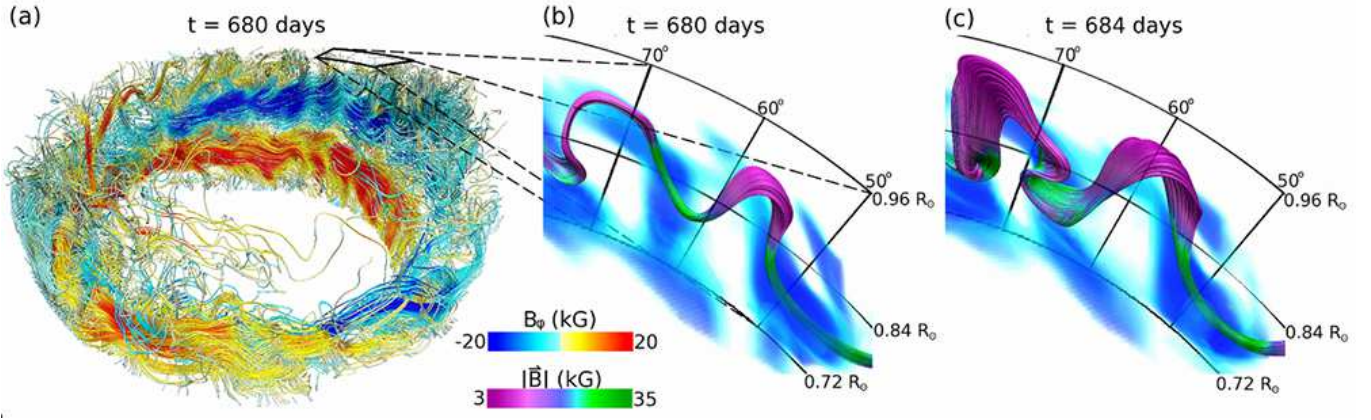}
 \caption{Buoyant magnetic loops evolving from small-scale wreath sections amplified by turbulent intermittency.  (a)~Field line rendering of magnetic wreaths at low latitudes in case S3. Field lines are colored by $B_\phi$ (negative in blue, positive in red) to highlight the two wreaths present. (b)~Zoom-in on region indicated in (a) showing field line tracings of the core of the buoyant magnetic loops at the same instant colored by magnitude of $\vec{B}$ (weak fields in purple, intense fields in yellow).  Volume rendering shows $B_\phi$ using the same color scheme as in (a). (c) The same region 4 days later, showing the continued rise of the loops through the stratified domain and their expansion.
 \label{fig:Loops}}
\end{center}
\end{figure*}

If these superequipartition wreath cores form adiabatically, this pressure deficit will be accompanied by a density deficit $\epsilon = \delta \rho/\rho \sim \delta P / (\gamma P)$, established by diverging flows along the axis of the wreath core. 
Radiative heating can further warm and rarify the wreath cores, enhancing the the density deficit to  $\epsilon \sim \delta P / P = (P_m - P_t)/P$ on a time scale of 
\begin{equation} \tau_r^{-1} = \frac{\epsilon}{r^2 \rho T C_p} \frac{\partial}{\partial t} \left(r^2 \rho T C_P \kappa_r \frac{\partial T}{\partial r}\right)
\end{equation} 
where $\kappa_r$ is the radiative diffusivity \citep{Fan1996}.  
Inserting values from Model S \citep{Christensen1996} for $\epsilon \sim 10^{-6}$ yields $\tau_r < $ 100 days through most of the solar convection zone.  
This value of $\epsilon$ corresponds to an emergence time $\tau_e \sim (2 D/\epsilon g)$, of about 10-15 days, where $D$ is the depth of the convection zone, and a magnetic field strength of $B \sim (8 \pi \epsilon P)^{1/2} \sim $ 20-40 kG over and above the equipartition value.  

Convection can also promote the buoyant rise of a wreath segment  by introducing a finite-amplitude undular displacement, resulting in a draining of fluid from from the apex of the loop \citep{Jouve2009, Nelson2011, Weber2011}.  
This could in principle operate for any field strength but in practice weak fields will by shredded and reprocessed by convection before they emerge \citep[e.g.,][]{Fan2009}.

The dynamics discussed here are indeed exhibited by our most turbulent simulation, case S3.  
Relative to more diffusive simulations, this case generates more regions of strong, superequipartition fields, as demonstrated in Figure~\ref{fig:Bphi PDFs}, and these regions are located in coherent, intermittent  wreath cores, as illustrated in Figure~ \ref{fig:S3 Iso}.  
Figure~\ref{fig:Loops} highlights two examples in which such wreath cores become buoyant and rise.  
As discussed in \cite{Nelson2011} and \cite{Nelson2013}, the loops rise through the convection zone through the combined influence of magnetic buoyancy and advection, reaching as high as $0.94 R$ before they are dissipated by diffusion.  
The wreath which formed these two loops (and two others not shown) is not axisymmetric; rather, it spans about 180$^\circ$ in longitude, reaching peak field strengths of 45 kG.  
We expect the process to be even more efficient in stars where the intermittency is presumably much more extreme.  

In summary, this paradigm of turbulence-induced flux emergence postulates that the combined action of turbulent intermittency and rotational shear generates a broad distribution of toroidal magnetic structures and it is only the most extreme events, in the high-$B$ tail of the pdf, that become buoyant.  
It is analogous to the theory of turbulence-regulated star formation, whereby supersonic turbulence in interstellar molecular clouds generates a spectrum of density fluctuations but only the extreme events on the tail of the pdf are dense enough to trigger the Jeans instability and condense to form protostars \citep{Krumholz2005}.  
It is also closely related to the negative magnetic pressure instability described by \cite{Kleeorin1989}  \citep[see also][]{Rogachevskii2007,Kapyla2012,Kemel2012}, although it does not necessarily rely on the assumptions that underlie that instability analysis, namely scale separation, the invariance of the small-scale turbulent energy, and the proportionality between variations in the mean and turbulent magnetic energy (attributed to kinematic shredding).

The radial location of the flux bundles that ultimately form active regions depends on the kinetic energy density in the convection (FKE) relative to that in the differential rotation (DRKE), as well as the efficiency of magnetic pumping.  In the simulations presented here, DRKE/FKE $\gtrsim 1$, suggesting that the generation of the wreaths is efficient enough that they can persist in the convection zone despite magnetic pumping. 
If this ratio falls much below unity, as might be expected for lower rotation rates, the wreaths may get pushed toward the base of the convection zone.  
Likewise, if the simulations are over or under-estimating the efficiency of magnetic pumping, this will influence the location of flux generation and the threshold to trigger the magnetic buoyancy instability. 
However, the basic paradigm should still be valid.

The scenario outlined here may resolve several current observational and theoretical puzzles.  
In particular, the non-axisymmetric nature of turbulence-induced flux emergence is consistent with the results of \cite{Stenflo2012} who find that many large bipolar active regions on the Sun violate Hales's polarity rules, and furthermore, that the anti-Hale regions often occur at the same latitude as bipoles that obey Hale's rules.  
The fraction of anti-Hale magnetic regions increases from about 4\% for the largest active regions (flux $\Phi \gtrsim 10^{23}$ Mx) to more than 25\% for smaller bipoles with $\Phi \sim 10^{20}$ Mx.  
The result that more than 70\% of intermediate-sized bipoles ($\Phi \sim 10^{20}$ Mx) obey Hale's laws suggests the presence of organized toroidal flux systems throughout the convection zone since all of these regions are unlikely to be anchored in the tachocline.  
Meanwhile, the diminishing of magnetic activity patterns with decreasing flux, including an increasing
fraction of anti-Hale bipoles as well as an increased scatter in tilt angles and emergence latitudes, is often attributed to the influence of convection on rising flux tubes \citep{Jouve2009, Weber2011, Weber2012, Jouve2012}.  
We propose that this intimate coupling between flux tubes and convection exists not only in their rise, but also in their very formation.  
Finally, the non-axisymmetric nature of turbulence-induced flux emergence may also account for the phenomenon of active longitudes \citep{Nelson2013}.

The observed tilt angles and emergence latitudes of bipolar magnetic regions on the Sun is best reproduced by models of rising flux tubes with initial field strengths of 20-100kG \citep[e.g.,][]{Fan2009, Jouve2009, Weber2011, Pinto2011}.  
However, generating such super-equipartition fields is not a trivial matter and in fact represents a formidable, unresolved problem in solar dynamo theory \citep[e.g.,][]{Rempel2001}.
Laminar amplification of toroidal fields by rotational shear, the $\Omega$-effect, tends to saturate at field strengths well below equipartition due to the back-reaction of the Lorentz force \citep{Vasil2009, Guerrero2011}.  
Turbulent intermittency can help by tapping the energy in the convection that is ultimately provided by the solar luminosity.  
It is clear from Figure~\ref{fig:Bphi PDFs} that the coupled action of turbulence and shear can generate superequipartition fields of the required amplitude.

The paradigm proposed here may also help address other difficulties with tachocline-based dynamos discussed by \cite{Brandenburg2005}. 
For example, toroidal flux generation does not rely on the radial shear of the tachocline, which is maximum near the poles. 
Instead, the expected location of flux generation is where $\vert \nabla \Omega\vert$ is maximum in the convection zone. 
This corresponds to the latitudinal shear at mid-latitudes, precisely where active regions first emerge at the beginning of a cycle, as emphasized by \cite{Spruit2010}.  
Note that the potential role flux emergence plays in establishing the solar cycle is a separate
question that we do not address here.

\section{Richness of stellar dynamos}

In this paper we have explored the complex behavior of a class of numerical simulations of convective dynamo action in rapidly rotating solar-like stars.  
More broadly, however, we have also touched upon the rich landscape of convective dynamo simulations by discussing both persistent wreath-building dynamos such as cases D3 and D3-pm1, and cyclic wreath-building dynamos including cases D3a, D3b, D3-pm2, and S3.  
Although the simulations considered here are ostensibly rotating three times faster than the Sun ($3 \Omega_\odot$), the Sun may actually be in a similar Rossby number regime, as noted in \S1. 
Thus the results presented here may have some bearing on the solar dynamo as well as the dynamos of younger, more rapidly-rotating solar analogues.

We have focused on two open questions that arose out of our previous work on wreath-building dynamos.  
The first is ``Can magnetic wreaths persist in the highly turbulent conditions of a stellar convection zone" and the second is ``What physical mechanisms establish and regulate the magnetic cycles we see in our simulations?"  
We have also touched upon a third question that all solar and stellar dynamo models must eventually face, and that is ``How are sunspots and bipolar active regions produced from dynamo-generated magnetic fields?".  

The principal issue with regard to the first question is whether magnetic wreaths can persist in stellar convection zones where the magnetic, viscous, and thermal diffusion coefficients are many orders of magnitude lower than in our simulations.  
We have investigated this question by systematically decreasing the diffusion in our simulations along two complementary paths, one in which only the magnetic diffusion coefficient, $\eta$, was reduced, and one in which the magnetic and viscous diffusivitiy, $\eta$ and $\nu$, were reduced together, keeping the magnetic Prandtl number constant (at a value of 0.5).  
In both cases magnetic wreaths with quasi-cyclic polarity reversals were attained, although the constant Pm branch exhibited more regular spatial and temporal behavior and thus became the focus of our analysis (see Figure~\ref{fig:TimeLat Stack}). 

Although no simulation can approach the extreme parameter regimes of stellar interiors, we have demonstrated a shift in the dynamical balances that bodes well for the possible persistence of magnetic wreaths at much higher Reynolds and magnetic Reynolds numbers.  
In short, our simulations suggest that the answer to the first question may be ``Yes, magnetic wreaths may indeed occur in actual stars".  
We have investigated in particular the balance of angular momentum transport which maintains the differential rotation in our simulations and the balance of processes responsible for creating and destroying the magnetic energy of the wreaths.  
In both cases, as we move from case D3 to case D3b we find that resolved turbulent dissipation has taken the place of SGS dissipation (see Figures~\ref{fig:DRKE Balances} and \ref{fig:Mag Energy Production}).  
This is an important milestone towards demonstrating that wreaths can exist in highly turbulent settings and that they are not reliant on the explicit diffusion in previous simulations.

We have found that magnetic wreaths persist in our higher-resolution, lower-dissipation, more turbulent simulations, yet their nature is altered in a fundamental and significant way.  
Most notably, they are no longer axisymmetric.  
In our more turbulent simulations such as case D3b, the nearly axisymmetric wreaths of case D3 are replaced by coherent wreath segments, typically spanning between 45$^\circ$ and 270$^\circ$ in longitude.  
This is associated with a shift in the magnetic power spectrum from longitudinal wavenumber $m = 0$ to moderate $m$ values.  
It also has important implications for flux emergence, as discussed with regard to question 3 below.

The first clues as to the origin of magnetic cycles in our simulations (question 2) were uncovered by \cite{Brown2010, Brown2011}, showing that one can move from a persistent wreath-building dynamo state to a cyclic one by increasing the rotation rate.  
Here we have shown that a similar transition from persistent to cyclic wreaths can be achieved by decreasing the effective magnetic diffusion, and thereby increasing the magnetic Reynolds number at a fixed rotation rate.  
As mentioned above, the constant-Pm branch of solutions exhibited more regular cyclic behavior despite the higher degree of turbulence.

We have not obtained a definitive exposition of the physical mechanisms that give rise to and regulate magnetic reversals.
However, we have traced their operation to the zonal component of the turbulent electromotive force (EMF) near the equator.  
In case D3 diffusion prevented reversals in the polarity of the axisymmetric poloidal field by locally offsetting the creation of mean poloidal field by turbulent fluctuations.  
In the lower-dissipation case D3b, this balance breaks down, leaving a residual turbulent EMF near the equator that creates poloidal field with a polarity that is opposite to that of the pre-existing field, as shown in Figure~\ref{fig:Aphi Balance D3b}.  
Once magnetic reversals are thus initiated, the overall reversal process follows the schematic description found in Figure~\ref{fig:Reversal Cartoon}.

Our simulations cannot directly address the third question regarding how solar and stellar dynamos produce sunspots and bipolar active regions.  
The detailed dynamics of flux emergence are too intricate to reliably capture in any current global dynamo simulation.  
However, the change in the nature of the wreaths as the dynamical balances shift suggests that they may play an important role in generating buoyant magnetic loops in actual stars.  
As discussed in \S7, these simulations suggest that strong, coherent magnetic structures of moderate angular extent can be created in the cores of the magnetic wreaths.
If this trend were to continue to the extremely low diffusion regimes of actual stellar convection zones, one would expect these flux bundles to become buoyantly unstable and rise.  
Indeed, this expectation is confirmed by our simulation Case S3 that employs a less diffusive SGS model and that exhibits the self-consistent generation of buoyant toroidal flux tubes in a global convective dynamo simulation.  
This picture of flux emergence as a fundamentally turbulent process contrasts strongly with more idealized scenarios where the principal role of convection is simply to produce a differential rotation.  
One might expect this revised paradigm to have observable consequences in such active region characteristics as distribution, tilt angle, and helicity.   
Furthermore, it may call into question our traditional reliance on sunspots as a straightforward proxy for the axisymmetric toroidal field at or below the base of the convection zone.

The rich behavior of these systems provides important insight into the dynamo models for the Sun and solar-type stars.  
The trend towards non-axisymmetric fields with enhanced turbulence, while still maintaining global-scale organization, pushes at the boundaries of our understanding of dynamo theory in solar-like settings.  
That these mechanisms are accessible with current computational resources clearly invites further intensive study of these topics.

\acknowledgements
We thank Kyle Auguston, Chris Chronopoulos, Yuhong Fan, Nicholas Featherstone, Brandley Hindman, Matthias Rempel, and Ellen Zweibel for their suggestions and advice.  This research is partly supported by NASA through
Heliophysics Theory Program grants NNX08AI57G and NNX11AJ36G.
Brown is supported through NSF Astronomy and Astrophysics postdoctoral
fellowship AST 09-02004.  CMSO is supported by NSF grant PHY 08-21899.
Miesch is also supported by NASA SR\&T grant NNH09AK14I.  NCAR is
sponsored by the National Science Foundation. Brun is partly
supported by both the Programmes Nationaux Soleil-Terre and Physique
Stellaire of CNRS/INSU (France), and by the STARS2 grant 207430 from
the European Research Council.  The simulations were carried out with
NSF TeraGrid and XSEDE support of Ranger at TACC, and Kraken at NICS, and with NASA HECC support of Pleiades.  
Field line tracings and volume renderings shown in Figures~\ref{fig:meet D3 family}, \ref{fig:S3 Iso}, and \ref{fig:Loops} were produced using VAPOR \citep{Clyne2007}.

\begin{appendix}

\section{The Dynamic Smagorinsky SGS Model in ASH}

We implement the dynamic Smagorinsky sub grid-scale model in the ASH code following the prescription of \cite{Germano1991}, inspired on the original formulation by \cite{Smagorinsky1963}. The Smagorinsky eddy viscosity $\nu_S$ is defined as
\begin{equation}
\nu_S = C_S \Delta^2 \left( \hat{e}_{ij} \hat{e}_{ij} \right)^{1/2} ,
\label{eq:DS}
\end{equation}
where $\hat{e}_{ij}$ is the resolved strain-rate tensor, $\Delta$ is the grid spacing, and $C_S$ is the constant of proportionality.
The dynamic Smagorinsky model (used here in case S3) makes an assumption of self-similar behavior in a resolved inertial range of a turbulent cascade to determine the value of $C_S$ at each point in the domain.  This is done by choosing a test-scale which is larger than the grid-scale by a factor $\beta$. Traditionally, and in this work, $\beta = 2$.  Only high resolution ASH simulations are able to assure that $\beta \Delta$ is suitably within an inertial range.

If we define the full velocity field, including scales that are not resolved in the simulation, as $u_i$, then the hat operator denotes grid-scale filtering keeping only the scales that are resolved in the simulation and the tilde operator expresses filtering at the test scale keeping only scales that are resolved by the simulation and larger that $\beta \Delta$. We define the stress tensor at the grid-scale $\Delta$ as 
\begin{equation}
{\bf \hat{S}}_{ij} = \widehat{u_i u_j} - \hat{u_i} \hat{u_j}
\end{equation}
and the stress tensor at the test filter scale $\beta \Delta$ as
\begin{equation}
{\bf \tilde{\hat{S}}}_{ij} = \widetilde{\widehat{u_i u_j}} - \tilde{\hat{u_i}} \tilde{\hat{u_j}} .
\end{equation}
Finally, the resolved stresses can be written as 
\begin{equation}
{\bf \cal{S}}_{ij} = \widetilde{\hat{u_i} \hat{u_j}} - \tilde{\hat{u_i}} \tilde{\hat{u_j}} .
\label{eqn:calS}
\end{equation}
Note that ${\bf \cal{S}}_{ij}$ can be computed directly from the resolved flows while ${\bf \hat{S}}_{ij}$ and ${\bf \tilde{\hat{S}}}_{ij}$ require a SGS model, such as the Smagorinsky model.

By assuming scale-invariance in the inertial range of the turbulent spectra, one obtains 
\begin{equation}
 {\bf \cal{S}}_{ij} = {\bf \tilde{\hat{S}}}_{ij}  - {\bf \hat{S}}_{ij} .
 \end{equation}
Note that this is simply stating that the resolved stresses are the difference between the stresses filtered at the test scale and the stresses filtered at the grid scale.  Applying the Smagorinsky model and contracting with ${\bf \hat{e}}_{ij}$ to obtain a scalar equation results in 
\begin{equation}
{\bf \cal{S}}_{ij} {\bf \hat{e}}_{ij} = -2 C_S \left( \beta^2 \Delta^2 \left| {\bf \hat{\tilde{e}}}_{ij} \right| {\bf \hat{\tilde{e}}}_{ij} {\bf \hat{e}}_{ij} - \Delta^2 \left| {\bf \hat{e}}_{ij} \right| {\bf \hat{e}}_{ij} {\bf \hat{e}}_{ij} \right) .
\end{equation}
Solving for $C_S$ gives
\begin{equation}
 C_S = - \frac{ {\bf \cal{S}}_{ij} {\bf \hat{e}}_{ij} }{ 2 \Delta^2 {\bf \hat{e}}_{ij} \left( \beta^2 \left| {\bf \hat{\tilde{e}}}_{ij} \right| {\bf \hat{\tilde{e}}}_{ij} - \left| {\bf \hat{e}}_{ij} \right| {\bf \hat{e}}_{ij}  \right) } .
 \label{eqn:c_s}
\end{equation}
To assure computational stability we require $C_S$ to be positive and apply a spectral filter on the denominator of Equation~(\ref{eqn:c_s}) which removes scales smaller than $\beta \Delta$. Additionally a Gaussian smoothing with a width equal to the largest horizontal grid spacing is applied to $C_S$ to prevent grid-scale ringing in the $C_S$ field. With the SGS viscosity thus determined, we apply constant SGS Prandtl and magnetic Prandtl numbers in order to determine the SGS thermal diffusivity and magnetic resistivity coefficients at each point in the domain.

The dynamic procedure involving a turbulent cascade has known problems adjacent to impenetrable walls where viscous boundary layers form that are rather laminar, such as the upper and lower boundaries in ASH \citep{Meneveau2000}. 
To compensate for a diminished cascade, we introduce a smoothly varying enhanced eddy viscosity immediately adjacent to there boundaries, occupying only 0.3\% of the domain in radial extent.

In the dynamic Smagorinsky model the nonlinear nature of the diffusion term requires an explicit time stepping scheme which imposes an upper limit on the size of our time-step. In order to control the time-step, an artificial ceiling is placed on the dynamic Smagorinsky viscosity. In case S3, on average, the ceiling is applied to 800 out of 76 million grid points at each time step.
As the time step is required to be less than $\Delta_\text{min}^2 / \nu$ for numerical stability, the functional form of the ceiling is given by 
\begin{equation}
\nu_{\text{max}} = t_{\text{max}} \Delta_{\text{min}}^2 ,
\end{equation}
where $\Delta_\text{min}$ is the smallest local grid-spacing in any direction and $t_{\text{max}}$ is the desired size time step.  In case S3 $t_{\text{max}}$ is set to 125 seconds.

\section{Generation of Differential Rotation Kinetic Energy}

As shown in Equation~(\ref{eq:Amom1}), the time evolution of angular momentum in our domain can be written in conservative form as the divergence of a flux vector $\vec{\mathcal{F}}$. The radial component is given by 
%
\begin{align}
\mathcal{F}_r =  \bar{\rho} \lambda & \left[ - \nu r \frac{ \partial }{ \partial r } \left( \frac{ v_\phi }{ r } \right) +  \widehat{v^\prime_\phi v^\prime_r} + \hat{v}_r \hat{v}_\phi + \hat{v}_r \Omega_0 \lambda \right. \nonumber \\
& \left. - \frac{1}{4 \pi \bar{\rho} } \widehat{B^\prime_\phi B^\prime_r}  - \frac{1}{4 \pi \bar{\rho} } \hat{B}_\phi \hat{B}_r  \right]  ,
\label{eq:AngMom1}
\end{align}
where the terms are from left to right due to viscous diffusion, fluctuating Reynolds stress, mean Reynolds stress from the meridional circulation, the Coriolis force with $\Omega_0$ representing the frame rotation rate, the Maxwell stress, and mean magnetic torques.  The latitudinal component is given by
\begin{align}
\mathcal{F}_\theta = \bar{\rho} \lambda & \left[ - \frac{ \nu \sin{\theta} } { r } \frac{ \partial }{ \partial \theta } \left( \frac{ v_\phi }{ \sin{\theta} } \right) +  \widehat{v^\prime_\phi v^\prime_\theta} + \hat{v}_\theta \hat{v}_\phi + \hat{v}_\theta \Omega_0 \lambda \right. \nonumber \\
& \left. - \frac{1}{4 \pi \bar{\rho} } \widehat{B^\prime_\phi B^\prime_\theta}  - \frac{1}{4 \pi \bar{\rho} } \hat{B}_\phi \hat{B}_\theta  \right] 
\label{eq:AngMom2}
\end{align}
where the terms have the same ordering and identities as in the radial component.  We ignore the flux due to the Coriolis force because while it can be large locally, it cannot do any net work on the system when averaged over the full domain. We can also write the fluxes in cylindrical coordinates in terms of the cylindrical radius $\lambda$ and the distance from the equatorial plane $z$. The flux in cylindrical radius is given by
\begin{equation}
\mathcal{F}_\lambda \hat{\lambda} = F_r \sin{\theta} \hat{r} + F_\theta \cos{\theta} \hat{\theta} ,
\end{equation}
while the flux in $z$ is given by
\begin{equation}
\mathcal{F}_z \hat{z} = F_r \cos{\theta} \hat{r} - F_\theta \sin{\theta} \hat{\theta} ,
\end{equation}

If we multiply equation (\ref{eq:Amom1}) by the longitude-averaged rotation profile $\hat{\Omega}$, we are left with an equation for the time evolution of the kinetic energy density in the mean differential rotation profile $\langle E_{\mathrm{DR}} \rangle$, 
  \begin{equation}
  \frac{\partial \langle E_{\mathrm{DR}} \rangle }{ \partial t } = \hat{\Omega} \left( \nabla \cdot \vec{\mathcal{F}} \right) .
  \end{equation}
We take a volume integral over the entire domain in order to calculate the total rate of change in the kinetic energy of differential rotation and rewrite the right-hand side as
  \begin{equation}
 \int_\mathcal{V} \frac{\partial \langle E_{\mathrm{DR}} \rangle }{ \partial t } dV = \int_\mathcal{V} \left[ \vec{\mathcal{F}} \cdot \nabla \hat{\Omega} - \nabla \cdot \left( \hat{\Omega} \vec{\mathcal{F}} \right) \right] dV .
  \end{equation}
 The second term in the integral can be rewritten using the divergence theorem as a surface integral, leaving us with
  \begin{equation}
 \int_\mathcal{V} \frac{\partial \langle E_{\mathrm{DR}} \rangle }{ \partial t } dV =  \int_\mathcal{V} \vec{\mathcal{F}} \cdot \nabla \hat{\Omega} dV - \int_\mathcal{S} \hat{\Omega} \mathcal{F}_r  dS .
 \label{eq:DRKE boundary}
  \end{equation}
Our choice of impenetrable and stress-free boundaries causes all of the hydrodynamic terms terms in the surface integral to vanish on both the inner and outer boundaries.  Likewise our choice of a perfect conductor boundary condition on the lower surface causes both the fluctuating and mean magnetic torques to vanish there.  The choice of a potential field boundary condition on the upper surface forces the mean magnetic torques to be exactly zero, however it does in principle allow the Maxwell stress to be non-zero.  This reduces the surface integral to 
  \begin{equation}
-  \int_\mathcal{S} \hat{\Omega} \mathcal{F}_r  dS = \int^\pi_0 \int_0^{2 \pi} \frac{ \hat{\Omega} }{4 \pi } \widehat{B^\prime_\phi B^\prime_\theta} R_o^3 \sin^2{\theta} \, d \theta \, d \phi .
  \end{equation}
We have calculated this term to be about five orders of magnitude smaller than the volume integral term in Equation~(\ref{eq:DRKE boundary}) when averaged over long periods in cases D3, D3a, and D3b. We chose to  ignore this surface term in our analysis of time-averaged quantities.

The generation and dissipation of differential rotation kinetic energy can be written as the sum of five terms, as was done in Equation~(\ref{eq:DRKE_prod}).  Those terms, which represent viscous diffusion, Reynolds stress, meridional circulations, Maxwell stress, and mean magnetic torques, are given in turn by
\begin{align}
L_\mathrm{VD} = - \int_\mathcal{V} \bar{\rho} \nu  r \sin{\theta} & \left[ r \frac{ \partial }{ \partial r } \left( \frac{ v_\phi }{ r } \right) \frac{ \partial \hat{\Omega} }{\partial r} \right. \nonumber \\
& \left. + \frac{ \sin{\theta} } { r^2 } \frac{ \partial }{ \partial \theta } \left( \frac{ v_\phi }{ \sin{\theta} } \right) \frac{ \partial \hat{\Omega} }{\partial \theta}  \right] dV ,
\end{align}
\begin{equation}
L_\mathrm{RS} = \int_\mathcal{V} \bar{\rho} r \sin{\theta} \left[ \widehat{v^\prime_\phi v^\prime_r} \frac{ \partial \hat{\Omega} }{\partial r}  + \frac{1}{r} \widehat{v^\prime_\phi v^\prime_\theta} \frac{ \partial \hat{\Omega} }{\partial \theta} \right] dV ,
\end{equation}
\begin{equation}
L_\mathrm{MC} = \int_\mathcal{V} \bar{\rho}r \sin{\theta} \hat{v}_\phi \left[ \hat{v}_r \frac{ \partial \hat{\Omega} }{\partial r} + \frac{ \hat{v}_\theta }{r} \frac{ \partial \hat{\Omega} }{\partial \theta} \right] dV ,
\end{equation}
\begin{equation}
L_\mathrm{MS} = - \int_\mathcal{V} \frac{ r \sin{\theta} }{ 4 \pi } \left[ \widehat{B^\prime_\phi B^\prime_r} \frac{ \partial \hat{\Omega} }{\partial r} + \frac{ \widehat{B^\prime_\phi B^\prime_\theta} }{r} \frac{ \partial \hat{\Omega} }{\partial \theta} \right] dV ,
\end{equation}
\begin{equation}
L_\mathrm{MT} = - \int_\mathcal{V} \frac{ r \sin{\theta} \hat{B}_\phi }{ 4 \pi } \left[ \hat{B}_r \frac{ \partial \hat{\Omega} }{\partial r} + \frac{ \hat{B}_\theta }{r} \frac{ \partial \hat{\Omega} }{\partial \theta} \right] dV .
\end{equation}
The time-averaged values of these terms are reported in Table~\ref{table:DRKE} and Figure~\ref{fig:DRKE Balances}.

\end{appendix}

\bibliographystyle{apj}
\bibliography{3SolarCycles}

\end{document}